\documentclass{aa}

\usepackage[utf8]{inputenc}
\usepackage[T1]{fontenc}

\usepackage[english]{babel}
\usepackage{txfonts}
\usepackage{charter}
\usepackage{balance}
\usepackage{hyperref}
    \hypersetup{colorlinks=true,linkcolor=blue,citecolor=blue,urlcolor=blue}
    \makeatletter
    \renewcommand*\aa@pageof{, page \thepage{} of \pageref*{LastPage}}
    \makeatother

\usepackage{mathabx}
\usepackage{siunitx}
    \DeclareSIUnit\angstrom{\text{Å}}
    \DeclareSIUnit\au{\text{AU}}

\usepackage{booktabs}
\usepackage[flushleft,para]{threeparttable}
\usepackage{multirow}
\usepackage{scrextend}

\usepackage[nowarn]{glossaries}
    \makeglossaries

\newacronym{1D}{1D}{one-dimensional}

\newacronym{API}{API}{application programming interface}
\newacronym{ARVE}{ARVE}{Analyzing Radial Velocity Elements}
\newacronym{AU}{AU}{astronomical unit}

\newacronym{BERV}{BERV}{barycentric Earth RV}
\newacronym{BIS}{BIS}{bisector inverse slope}

\newacronym{CCF}{CCF}{cross-correlation function}
\newacronym{CSV}{CSV}{comma-separated values}

\newacronym{DACE}{DACE}{Data \& Analysis Center for Exoplanets}
\newacronym{DRP}{DRP}{data reduction pipeline}
\newacronym{DRS}{DRS}{data reduction software}

\newacronym{EPRV}{EPRV}{extreme precision RV}

\newacronym{FAP}{FAP}{false-alarm probability}
\newacronym{FITS}{FITS}{Flexible Image Transport System}
\newacronym{FWHM}{FWHM}{full-width at half-maximum}

\newacronym{GLS}{GLS}{generalized Lomb-Scargle}
\newacronym{GP}{GP}{Gaussian process}

\newacronym{HITRAN}{HITRAN}{High-resolution Transmission Molecular Absorption Database}
\newacronym{HWHM}{HWHM}{half-width at half-maximum}

\newacronym{LBL}{LBL}{line-by-line}
\newacronym{LTE}{LTE}{local thermodynamic equilibrium}

\newacronym{MARCS}{MARCS}{Model Atmospheres with a Radiative and Convective Scheme}

\newacronym{NaN}{NaN}{Not a Number}
\newacronym{NIR}{NIR}{near-infrared}

\newacronym{PBP}{PBP}{part-by-part}

\newacronym{RMS}{RMS}{root-mean-square}
\newacronym{RV}{RV}{radial velocity}

\newacronym{S/N}{S/N}{signal-to-noise ratio}
\newacronym{SIMBAD}{SIMBAD}{Set of Identifications, Measurements and Bibliography for Astronomical Data}
\newacronym{SME}{SME}{Spectroscopy Made Easy}

\newacronym{VALD}{VALD}{Vienna Atomic Line Database}
\newacronym{VPSD}{VPSD}{velocity power spectral density}

\usepackage{orcidlink}

\begin{document}

    \title{ARVE: Analyzing Radial Velocity Elements}
    \subtitle{I. The Code}

    \author{K. Al Moulla$^{1,2,}$\thanks{SNSF Postdoctoral Fellow} \orcidlink{0000-0002-3212-5778}}

    \institute{Instituto de Astrofísica e Ciências do Espaço, Universidade do Porto, CAUP, Rua das Estrelas, 4150-762 Porto, Portugal\\
    \email{khaled.almoulla@astro.up.pt}
    \and{Observatoire Astronomique de l'Université de Genève, Chemin Pegasi 51, 1290 Versoix, Switzerland}}

   \date{Received 31 March 2025 / Accepted 22 July 2025}

\abstract
{In order to overcome the \acrfull*{RV} precision barrier imposed by stellar variability, there has in recent times been a surge of software aimed at simulating and modeling different aspects of these activity patterns which currently limit the feasibility of detecting Earth-like exoplanets.}
{We present \acrlong*{ARVE} (\texttt{ARVE}), a Python-based software which enables RV extraction using various customizable techniques, and subsequent analysis of the stellar and planetary signals present in the RVs. One of \texttt{ARVE}'s unique features is its library of pre-computed auxiliary data, which includes synthetic spectra and spectral line masks, allowing the code to efficiently perform certain routines with minimal input from the user.}
{\texttt{ARVE} is a class-based and modular code in which its functionalities are divided between four subclasses: \texttt{functions}, which handles general functions utilized by the other subclasses; \texttt{data}, which reads the input data, loads the auxiliary data, and extracts RVs from input high-resolution spectra; \texttt{star}, which characterizes the stellar activity components present in the RV time series; and \texttt{planets}, which performs fits of Keplerian signals in the data and offers injection-recovery tests of fictitious planets to determine the detection limits.}
{Demonstrations of \texttt{ARVE} are performed on three years of HARPS-N solar data. We show the evolution of granulation and supergranulation characteristic timescales with activity level, and we investigate the differences in planetary period-mass detection limits when extracting RVs with different methods.}
{As stellar activity mitigation techniques grow more diverse, we foresee that a tool like \texttt{ARVE} could greatly benefit the community by offering a user-friendly and multi-functional approach to extract and analyze RV time series. With its current code structure, expanded functionality and increased compatibility with more spectrographs should be easily addable to future versions of \texttt{ARVE}.}


\authorrunning{K. Al Moulla}
\titlerunning{ARVE: Analyzing Radial Velocity Elements. I.}

\maketitle


\section{Introduction}\label{Sect:1}

As state-of-the-art high-resolution spectrographs have become able to reach \acrfull*{RV} precision below the meter-per-second level \citep{Pepe+2014}, one of the current primary obstacles in detecting and measuring the masses of Earth-analogous exoplanets---i.e., low-mass companions orbiting within the habitable zones of their host stars---with the RV method, is the pervasive presence of stellar variability manifesting as apparent Doppler shifts \citep{Crass+2021}. These include---in generally increasing characteristic timescales---oscillations induced by pressure waves \citep[e.g.,][]{Kjeldsen&Bedding1995}, granulation phenomena caused by turbulent convective motions \citep[e.g.,][]{Meunier+2015}, stellar surface regions formed by concentrated magnetic fields \citep[e.g.,][]{Saar&Donahue1997,Meunier+2010}, and magnetic cycles which dictate the coverage of the aforementioned surface regions \citep[e.g.,][]{Haywood+2022}. The RV variations which they produce can be $1$--$2$ orders of magnitude larger than the amplitudes of Earth twins (which is around $\SI{10}{\centi\meter\per\second}$), and can moreover have a quasi-periodic nature mimicking true planetary signals \citep[e.g.,][]{Lubin+2021,Carmona+2023}. Therefore, it has become increasingly urgent to understand and mitigate the influence of these stellar activity patterns, to make \acrfull*{EPRV} feasible, and to ensure the success of future missions, e.g., PLATO \citep{Rauer+2014}, aiming to detect and characterize terrestrial planets.

Recently, a surge of stellar activity-targeting software has enabled the simulation and measurement of variability constituents in RVs (and notably also photometry). For example, the \texttt{GRASS} code \citep{Palumbo+2022,Palumbo+2024} simulates spectral line asymmetries caused by granulation, and the \texttt{SOAP} code \citep{Boisse+2012} and its adaptations \citep{Oshagh+2013,Dumusque+2014,Zhao&Dumusque2023} simulate the impact of magnetically active surface regions. Codes which enable the measurement of the activity contribution in RVs include approaches using translationally separable representations of the stellar spectrum as in $\phi$\texttt{ESTA} \citep{Zhao&Tinney2020,Zhao+2022} and \texttt{SCALPELS} \citep{CollierCameron+2021}, principal component analysis of spectral line behaviors as in \texttt{YARARA} \citep{Cretignier+2021,Cretignier+2023} and \texttt{Wapiti} \citep{Ould-Elhkim+2023}, \acrfull*{GP} regression as in \texttt{PYANETI} \citep{Barragan+2019,Barragan+2022} and \texttt{MAGPy-RV} \citep{Rescigno+2023}, and machine learning as in \texttt{CALM} \citep{deBeurs+2024} and \texttt{AESTRA} \citep{Liang+2024}.

In this paper, we present \acrlong*{ARVE}, abbreviated and stylized as \texttt{ARVE}, a novel Python-based software, which enables the extraction of RVs from high-resolution stellar spectra with different, customizable methods, and their subsequent analysis in terms of stellar activity characterization and planetary signal detection.

The following sections are organized as follows. In Sect.~\ref{Sect:2}, we describe the details and purpose of the auxiliary data which is generated beforehand and included alongside the code. In Sect.~\ref{Sect:3}, we describe the structure of the code, and the mathematical framework and numerical implementation of its functionalities. In Sect.~\ref{Sect:4}, we demonstrate some of the code's capabilities on solar data. Finally, in Sect.~\ref{Sect:5}, we discuss the possible role of the code in future endeavors and how it could be eventually expanded.

\section{Auxiliary data}\label{Sect:2}

To increase its computational efficiency, \texttt{ARVE} utilizes pre-computed auxiliary data complementary to the user input. Parts of the auxiliary data (the telluric spectrum and spectral line masks; see Sects.~\ref{Sect:2.2} and \ref{Sect:2.3}, respectively) are directly included with the package at installation, whereas the remaining data products (the stellar spectra; see Sect.~\ref{Sect:2.1}) are downloaded the first time they are required by the code. The reason for this partition is the current $\SI{60}{MB}$ size limit on Python packages distributed through the PyPI repository.

Including this auxiliary data, rather than generating it on the fly, offers three key advantages: (1) improved computational speed, as loading pre-generated data is significantly faster than synthesizing it; (2) reduced dependencies, as \texttt{ARVE} can function independently of any specific synthesis code (with the option to upgrade included products in the future); and (3) valuable resources for external use, such as quick-look spectral content across different spectral types and ready-to-use line masks.

\subsection{Stellar spectra}\label{Sect:2.1}

\begin{table*}[t!]
	\caption{Stellar parameters for the grid of spectral types provided in the auxiliary data (see Sect.~\ref{Sect:2}).}
	\begin{tabular*}{\textwidth}{l @{\extracolsep{\fill}} S[table-format=4.0] @{\extracolsep{\fill}} S[table-format=1.1] @{\extracolsep{\fill}} S[table-format=1.2] @{\extracolsep{\fill}} S[table-format=1.2] @{\extracolsep{\fill}} S[table-format=3.1]}
		\toprule
		\midrule
		{Sp. type} & {$T_{\mathrm{eff}}$ [$\SI{}{\kelvin}$]} & {$\log{g}$ [cgs]} & {$M$ [$M_{\Sun}$]} & {$R$ [$R_{\Sun}$]} & {$v\sin{i}$ [$\SI{}{\kilo\meter\per\second}$]} \\
		\midrule
		F0 & 7178 & 4.3 & 1.66 & 1.62 & 180.0 \\
		F2 & 6909 & 4.3 & 1.56 & 1.48 & 135.0 \\
		F5 & 6528 & 4.3 & 1.41 & 1.40 &  20.0 \\
		F8 & 6160 & 4.4 & 1.25 & 1.20 &   9.0 \\
		G0 & 5943 & 4.4 & 1.16 & 1.12 &   6.4 \\
		G2 & 5811 & 4.4 & 1.11 & 1.08 &   4.8 \\
		G5 & 5657 & 4.5 & 1.05 & 0.95 &   3.4 \\
		G8 & 5486 & 4.5 & 0.97 & 0.91 &   2.6 \\
		K0 & 5282 & 4.6 & 0.90 & 0.83 &   2.2 \\
		K2 & 5055 & 4.6 & 0.81 & 0.75 &   2.0 \\
		K3 & 4973 & 4.6 & 0.79 & 0.73 &   2.0 \\
		K5 & 4623 & 4.6 & 0.65 & 0.64 &   1.9 \\
		K7 & 4380 & 4.7 & 0.54 & 0.54 &   1.7 \\
		M0 & 4212 & 4.7 & 0.46 & 0.48 &   1.5 \\
		M2 & 4076 & 4.7 & 0.40 & 0.43 &   0.0 \\
		M5 & 3923 & 4.8 & 0.34 & 0.38 &   0.0 \\
		\bottomrule
	\end{tabular*}
	\footnotesize
	The columns specify the spectral type, effective temperature, $T_{\mathrm{eff}}$, in Kelvin, logarithmic surface gravity, $\log{g}$, in cgs units, mass, $M$, in solar masses, radius, $R$, in solar radii, and equatorial rotational velocity, $v\sin{i}$, in kilometers per second.
	\label{Tab:01}
\end{table*}

In order to have \textit{a priori} knowledge about the stellar line properties of different spectral types, \texttt{ARVE} makes use of a grid of pre-computed synthetic spectra. These are subsequently used for multiple purposes, including the generation of spectral line masks (see Sect.~\ref{Sect:2.3}), and the RV extraction with various methods (see Sects.~\ref{Sect:3.2.3} and \ref{Sect:3.2.4}).

The grid consists of $16$ main-sequence spectral types, ranging from F0 to M5. Their stellar parameters are listed in Table~\ref{Tab:01}, which is a subset of the values from Table B.1 in \cite{Gray2008}. To avoid including additional dimensions to the grid, all spectral types are assumed to have solar metallicities, where the abundances are taken from \cite{Asplund+2009}. For the choice of spectral synthesis, we selected \texttt{PySME} \citep{Wehrhahn+2023}, which is a Python adaptation of the \acrfull*{1D} \acrfull*{LTE} radiative transfer code \acrlong*{SME} \citep[\texttt{SME};][]{Valenti&Piskunov1996,Piskunov&Valenti2017}. The spectra are computed for the vacuum wavelengths $\SI{3000}{}$--$\SI{23\,000}{\angstrom}$, in steps of $\SI{0.01}{\angstrom}$, in order to cover the wavelength range used by most modern high-resolution optical and \acrfull*{NIR} spectrographs. For each spectral type we queried line lists from VALD3\footnote{VALD stands for \acrlong*{VALD}.\\Available at \url{http://vald.astro.uu.se}} \citep{Piskunov+1995,Kupka+2000,Ryabchikova+2015} in intervals of $\SI{1000}{\angstrom}$, with a minimum depth limit of $0.1$, to avoid reaching the $100\,000$ line threshold imposed by the VALD query system. The model atmospheres, interpolated by \texttt{PySME} at the specified stellar parameters, are chosen to be standard plane-parallel \texttt{MARCS}\footnote{\texttt{MARCS} stands for \acrlong*{MARCS}.\\Available at \url{https://marcs.astro.uu.se}} \citep{Gustafsson+2008} models. The final spectra were broadened with their rotational velocities, $v\sin{i}$, and their micro- and macroturbulences adopted from \cite{Valenti&Fischer2005}, where the authors fixed the microturbulence to
\begin{equation}\label{Eq:01}
    v_{\mathrm{mic}} = \SI{0.85}{\kilo\meter\per\second}\,,
\end{equation}
and scaled the macroturbulence with the effective temperature, $T_{\mathrm{eff}}$,
\begin{equation}\label{Eq:02}
    v_{\mathrm{mac}} = \left(3.98 + \frac{T_{\mathrm{eff}} - \SI{5770}{K}}{\SI{650}{K}}\right) \, \SI{}{\kilo\meter\per\second}\,.
\end{equation}
Due to an apparent typo in the equation in \cite{Valenti&Fischer2005}, we note that the sign of the $T_\mathrm{eff}$-term in Eq.~\ref{Eq:02} has been changed.

In addition to the normalized flux spectra, we also include the average formation temperature, denoted $T_{1/2}$ and defined as the photospheric temperature at which the cumulative flux contribution function is equal to half its maximum value \citep{AlMoulla+2022,AlMoulla+2024}, evaluated at each wavelength point of the syntheses.

A sample of the flux and formation temperature spectra are shown in Figs.~\ref{Fig:A01}--\ref{Fig:A05}.

\subsection{Telluric spectrum}\label{Sect:2.2}

For precise RV measurements where the stellar spectra have been wavelength-shifted to the stellar rest frame, it becomes important to mask out or correct the contamination from telluric molecular lines and bands. If unaccounted for, these tellurics will influence the measured RV by moving in and out the vicinity of stellar lines with the \acrfull*{BERV} which, depending on the orientation of the observed star relative to Earth, can be as high has as the sum of the Earth orbital and equatorial rotation velocities, i.e., ${\sim}\SI{30}{\kilo\meter\per\second}$. Since telluric correction is beyond the scope of the current functionalities of \texttt{ARVE}, the telluric spectrum included is a simplified model which includes the most prevalent molecular species at average atmospheric conditions, and is meant to be used only to mask out the most severely affected wavelength regions. For a more rigorous treatment of telluric correction, we refer the reader to more detailed telluric models which take the local meteorological conditions into account \citep[e.g.,][]{Allart+2022}.

The telluric spectrum was modeled using data from HITRAN\footnote{HITRAN stands for \acrlong*{HITRAN}.\\Available at \url{https://hitran.org}} \citep{Gordon+2022}. We used \texttt{HAPI} \citep{Kochanov+2016}, the HITRAN Python \acrfull*{API}, to fetch molecular line lists for the most dominant isotopes of water (H$_{2}$O), carbon dioxide (CO$_{2}$), methane (CH$_{4}$), and oxygen (O$_{2}$). The line lists were fetched between the wavenumbers, $\nu$, corresponding to the wavelength bounds of the synthetic stellar spectra (see Sect.~\ref{Sect:2.1}),
\begin{equation}\label{Eq:03}
    \nu = \frac{\SI[parse-numbers=false]{10^{8}}{\angstrom}}{\lambda} \, \SI{}{\per\centi\meter}\,,
\end{equation}
where $\lambda$ are the wavelengths given in units of \SI{}{\angstrom}. We then called the built-in functions in \texttt{HAPI} to compute the Lorentzian absorption coefficients of the aforementioned molecules in a typical air gas mixture. Finally, we used the absorption coefficients to compute the normalized transmittance spectrum, which was thereafter interpolated at the same wavelength grid as the synthetic stellar spectra.

\subsection{Spectral line masks}\label{Sect:2.3}

\begin{figure}[ht!]
    \includegraphics[width=\linewidth]{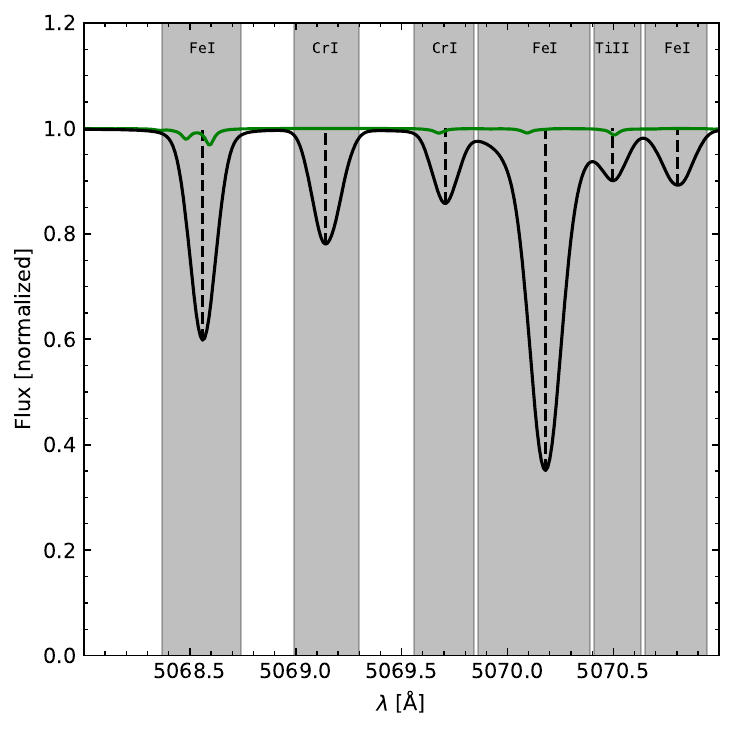}
    \caption{Example of spectral line mask portion for the G2 spectral type. The synthetic flux spectrum (black) and telluric spectrum (green) are shown together with the identified spectral lines (gray windows) whose wavelengths at flux minimum (black dashed lines) and elemental species (text annotations) are highlighted.}
    \label{Fig:01}
\end{figure}

The spectral line masks are tables with spectral line properties for each spectral type. Unlike the VALD line lists, the line masks also contain information about the line boundaries. For each spectral type, a line mask was generated by identifying local minima in the corresponding synthesized flux spectrum (see Sect.~\ref{Sect:2.1}). For each local minimum, the five closest points were fitted with a second-order polynomial in order to evaluate the central wavelength and depth. The spectral line with the closest wavelength from the corresponding VALD line list was taken as the identifier of that local minimum. For each spectral line, the lower and upper wavelength boundaries were identified as the points closest to the line center at which the normalized flux level reaches above $0.99$ or at which the flux gradient changes sign. The second condition ensures that blended lines are separated at their intersection. Additionally, the optimal weight per line, $w$, for RV extraction---hereafter referred to as the RV content---is also provided following \cite{Bouchy+2001},
\begin{equation}\label{Eq:04}
    w = \sum_{i}\left(\frac{\mathrm{d}F}{\mathrm{d}\lambda}\right)_{i}^{2}\frac{\lambda_{i}^{2}}{F_{i}}
\end{equation}
where $\lambda_{i}$ and $F_{i}$ are the wavelengths and fluxes, respectively, of the sampled points within the bounds of each line. A sample of the G2 spectral line mask is shown in Fig.~\ref{Fig:01}.

\section{The \texttt{arve} class}\label{Sect:3}

The \texttt{ARVE} code is a class-based and modularized software. Almost all functionality is handled and executed by creating a specialized class object which is thereafter used to call the class methods of the \texttt{arve} class and its subclasses. Note that we use uppercase \texttt{ARVE} to refer to the code as a whole, and lowercase \texttt{arve} to refer to the primary class.

The \texttt{arve} class has only five class variables: \texttt{id}, which is a label given to each instance of the class, and four subclasses, \texttt{functions}, \texttt{data}, \texttt{star}, and \texttt{planets}, which govern the different aspects of the code utility, and which are described in detail below.

In addition to the following functions, the subclasses have several plotting functions, all commencing with the prefix \texttt{plot\_}, which produce publication-ready figures of the various outputs. Their functionality is self-explanatory from the naming scheme, hence their description is omitted here, and we refer the reader to the official \texttt{ARVE} documentation for their usage.

\subsection{The \texttt{functions} subclass}\label{Sect:3.1}

The \texttt{functions} subclass is a helper class, containing functions which are required by one or several of the other subclasses, and is thus not intended to be called directly by the user. Its sole class variable is a dictionary with constants, alleviating the need to define them in several places.

The \texttt{functions} class variable is the following:
\begin{labeling}[~:]{\texttt{constants}}
	\item[\texttt{constants}] the constants
\end{labeling}

\subsubsection{\texttt{functions.convert\_vac\_to\_air()}}\label{Sect:3.1.1}

The \texttt{convert\_vac\_to\_air()} function handles vacuum to air conversion between wavelengths. Although most modern high-resolution spectrographs operate in vacuum, some of their pipelines output wavelength solutions in air medium. The conversion from vacuum wavelengths, $\lambda_{\mathrm{vac}}$, to air wavelengths, $\lambda_{\mathrm{air}}$, is given by
\begin{equation}\label{Eq:05}
    \lambda_{\mathrm{air}} = \frac{\lambda_{\mathrm{vac}}}{n} \,,
\end{equation}
where the chromatic solution of the refractive index, $n$, defined in \cite{Morton+2000}, is
\begin{equation}\label{Eq:06}
    n = 1.0000834254 + \frac{0.02406147}{130-s^{2}} + \frac{0.00015998}{38.9-s^{2}}
\end{equation}
and where $s\,{=}\,\SI[parse-numbers=false]{10^{4}}{\angstrom}\,/\,\lambda_{\mathrm{vac}}$.

\subsubsection{\texttt{functions.convert\_air\_to\_vac()}}\label{Sect:3.1.2}

The \texttt{convert\_vac\_to\_air()} function handles air to vacuum conversion between wavelengths. The conversion, similar to the inverse conversion above, is given by
\begin{equation}\label{Eq:07}
    \lambda_{\mathrm{vac}} = \lambda_{\mathrm{air}}n \,,
\end{equation}
where the refractive index is now adopted from a solution by N. Piskunov\footnote{Available at \url{https://www.astro.uu.se/valdwiki/Air-to-vacuum\%20conversion}},
\begin{equation}\label{Eq:08}
\begin{split}
    n &= 1.00008336624212083 + \frac{0.02408926869968}{130.1065924522-s^{2}} \\
      &+ \frac{0.0001599740894897}{38.92568793293-s^{2}}
\end{split}
\end{equation}
and where $s\,{=}\,\SI[parse-numbers=false]{10^{4}}{\angstrom}\,/\,\lambda_{\mathrm{air}}$.

\subsubsection{\texttt{functions.doppler\_shift()}}\label{Sect:3.1.3}

The \texttt{doppler\_shift()} function handles the Doppler wavelength shift due to the source of light moving with a velocity, $v$. The shifted wavelength, $\tilde{\lambda}$, as function of the wavelength at rest, $\lambda$, is given by
\begin{equation}\label{Eq:09}
    \tilde{\lambda} = \lambda \sqrt{\frac{1+v/c}{1-v/c}} \,,
\end{equation}
where $c\,{=}\,\SI{2.99792458e5}{\kilo\meter\per\second}$ is the vacuum speed of light. For non-relativistic velocities, the expression in Eq.~\ref{Eq:09} simplifies to
\begin{equation}\label{Eq:10}
    \tilde{\lambda} = \lambda(1+v/c) \,.
\end{equation}

\subsubsection{\texttt{functions.inverted\_gaussian()}}\label{Sect:3.1.4}

The \texttt{inverted\_gaussian()} function returns an inverted Gaussian, $\mathcal{IG}$, evaluated at specified abscissa points, $x$, of the following form,
\begin{equation}\label{Eq:11}
    \mathcal{IG}(x) = C\left( 1 - a\mathrm{exp}\left( -\frac{(x-b)^{2}}{2c^{2}} \right) \right) \,,
\end{equation}
where $C$ is the continuum level, $a$ is the normalized inverted Gaussian depth, $b$ is the $x$-coordinate of the central point, and $c$ is the standard deviation which relates to the \acrfull*{FWHM} through
\begin{equation}\label{Eq:12}
    \mathrm{FWHM} = 2\sqrt{2\log{2}}c \,.
\end{equation}

\subsubsection{\texttt{functions.gls\_periodogram()}}\label{Sect:3.1.5}

The \texttt{gls\_periodogram()} function returns the \acrlong*{GLS} \citep[GLS;][]{Lomb1976,Scargle1982,Zechmeister&Kurster2009} periodogram of a time series evaluated at the frequencies, $f$, bounded by the lowest resolvable frequency and the Nyquist frequency,
\begin{equation}\label{Eq:13}
    f = \frac{1}{T}, \frac{2}{T{\cdot}N}, \dots, \frac{1}{2{\Delta}t_{\mathrm{med}}} \,,
\end{equation}
where $T$ is the total time span, $N$ is the over-factorization (defaulting to 1), and ${\Delta}t_{\mathrm{med}}$ is the median time step. The periodogram finds the best-fitting linear combination of sinusoidal functions, including a constant offset, to the signal $y$ at times $t$ for each frequency,
\begin{equation}\label{Eq:14}
    y(t) = \sum_{f} a(f)\cos{2{\pi}ft} + b(f)\sin{2{\pi}ft} + c(f) \,,
\end{equation}
where $a$, $b$, and $c$ are the fitted coefficients. The periodogram can either be returned according to the normalization detailed in \cite{Zechmeister&Kurster2009}, or as a power spectrum, in squared physical units, given by $a^{2} + b^{2}$.

\subsection{The \texttt{data} subclass}\label{Sect:3.2}

The \texttt{data} subclass is responsible for the extraction, storage, and manipulation of all data which are not star- nor planet-specific, i.e., spectra, RV time series, instrument specifications, etc.

The \texttt{data} class variables are the following:
\begin{labeling}[~:]{\texttt{vrad\_components}}
    \item[\texttt{aux\_data}       ] the auxiliary data
	\item[\texttt{time}            ] the time stamps
    \item[\texttt{vrad}            ] the RVs
    \item[\texttt{vrad\_components}] the RV components
    \item[\texttt{spec}            ] the spectra
    \item[\texttt{spec\_reference} ] the reference spectrum
    \item[\texttt{ccf}             ] the CCFs (see Sect.~\ref{Sect:3.2.3})
\end{labeling}

\subsubsection{\texttt{data.add\_data()}}\label{Sect:3.2.1}

The \texttt{add\_data()} function handles the input data, which can either be an RV time series (for which time, RV, and RV error arrays need to be provided), or a spectral time series (for which wavelength, flux, and flux error matrices need to be provided, and optionally a time array).

For spectral time series, the user can either input the matrices themselves if they do not exceed the available RAM, or point to a directory where the spectra are stored as either \acrfull*{CSV}, \texttt{numpy} zipped archives (NPZ)\footnote{Available at \url{https://numpy.org/doc/2.2/reference/generated/numpy.lib.format.html}}, or \acrlong*{FITS} \citep[FITS;][]{Wells+1981} files. If supplied by the user directly, the spectral time series must be interpolated onto a common wavelength grid. The wavelength matrix must have dimensions $(N_{\mathrm{ord}},N_{\mathrm{pix}})$, where $N_{\mathrm{ord}}$ is the number of spectral orders, and $N_{\mathrm{pix}}$ is the number of pixels, and the flux matrices must have dimensions $(N_{\mathrm{spec}},N_{\mathrm{ord}},N_{\mathrm{pix}})$, where $N_{\mathrm{spec}}$ is the number of spectra. If stored externally, the spectra must be self-contained in individual files, however, they do not need to be interpolated onto the same the wavelength grid. If stored in CSV or NPZ format, the matrices must be called \texttt{wave\_val}, \texttt{flux\_val}, and \texttt{flux\_err}, respectively, in order to be recognized; if stored in FITS format, specifying the instrument and spectral format (i.e., if the spectra are echelle order-merged, known as S1D, or echelle order-separated, known as S2D\footnote{The naming conventions S1D and S2D, used by the pipelines of several high-resolution spectrographs, both refer to 1D spectra which have been reduced and wavelength calibrated. S1D denotes a spectrum where all the echelle orders have been merged together into one continuous spectrum, and S2D denotes a spectrum where the echelle orders are separated into different rows in the data matrix. In \texttt{ARVE}, S1D spectra are treated as order-separated spectra with $N_{\mathrm{ord}}\,{=}\,1$.}) is sufficient to read all required variables. As a reference spectrum, the first spectrum in the referenced directory is stored as a class variable. The input wavelengths are also corrected for the systematic stellar velocity (see Sect.~\ref{Sect:3.3.1}) upon loading/storing them.

Additionally, the user can optionally specify the spectral resolution (if the instrument has not been specified), whether the spectra are BERV corrected (and if not, a BERV array can be provided to perform the correction), and if they have a common wavelength grid (in order to circumvent the interpolation performed in some functions). For the interpolation, the default method is a cubic spline, although the user can select any of the available options in \texttt{scipy.interpolate.interp1d()}\footnote{Available at \url{https://docs.scipy.org/doc/scipy/reference/generated/scipy.interpolate.interp1d.html}}.

\subsubsection{\texttt{data.get\_aux\_data()}}\label{Sect:3.2.2}

The \texttt{get\_aux\_data()} function handles the fetching and adjustment of the auxiliary data complementary to the input data. Based on the stellar spectral type (which is either user-specified or fetched from a database; see Sect.~\ref{Sect:3.3.1}), the closest set of auxiliary data is fetched from the pre-computed grid summarized in Table~\ref{Tab:01}.

The synthetic stellar spectrum is shifted to the same medium as the observed wavelengths, broadened with a Gaussian instrumental profile corresponding to the resolution, and interpolated onto the observed wavelength grid.

The synthetic telluric spectrum undergoes the same treatment as the stellar, while additionally being shifted with the negative systematic velocity in order to align it with the observed spectra (which have also been corrected for the systematic motion). Thereafter, all telluric lines deeper than a user-specified threshold (defaults to $\SI{1}{\percent}$) are identified and shifted by a user-specified maximum BERV (defaults to $\pm\SI{1}{\kilo\meter\per\second}$ for the Sun, and $\pm\SI{30}{\kilo\meter\per\second}$ for other stars) in order to identify wavelength ranges affected by tellurics. Overlapping ranges are merged, such that the stored information is the lower and upper wavelength bounds of the smallest number of ranges.

The spectral line mask is also shifted to the same medium as the observations, and cropped to only contain lines within the relevant wavelength range. The pixel indices of the lower and upper line bounds are searched and stored for quick future access. Additionally, lines within the aforementioned telluric ranges are flagged to be optionally excluded in, e.g., the RV extraction functions.

\subsubsection{\texttt{data.compute\_vrad\_ccf()}}\label{Sect:3.2.3}

The \texttt{compute\_vrad\_ccf()} function extracts RVs from input spectra using the \acrlong*{CCF} \citep[CCF;][]{Baranne+1996,Pepe+2002}. In order to compute the CCF, a velocity grid over which the spectral line mask will be shifted, must be defined. If not provided by the user, the velocity range defaults to $[-20,20]\,\SI{}{\kilo\meter\per\second}$, with a step taken as the median pixel size converted to a velocity shift (around $\SI{1}{\kilo\meter\per\second}$ for modern high-resolution spectrographs). Next a line mask is constructed by selecting the lines from the auxiliary data (see Sect.~\ref{Sect:2.3}) which are within the observed wavelength range even when the lines are shifted to the minimal or maximal grid velocity. By default, all lines in the line mask are included, however, users may edit the line mask before the CCF computation step, or specify the application of certain criteria provided by the function.

To make the CCF calculation computationally efficient, the relevant wavelengths (or rather matrix indices for the numerical treatment) at which the spectra will be cross-correlated are identified and stored in a preceding step, and thereafter used for all spectra since they are (or will be) interpolated at the same wavelength grid. The relevant wavelengths are identified by shifting the central wavelength of each line in the line mask according the velocities in the velocity grid, and thereafter locating the closest observed wavelengths for each spectral order, spectral line, and velocity shift. $\lambda^{\mathrm{lower}}$ and $\lambda^{\mathrm{upper}}$ denotes the lower and upper closest observed wavelengths, respectively, and $r^{\mathrm{lower}}$ and $r^{\mathrm{upper}}$ denotes their respective fractional coverage of the mask line. Mask lines are assumed to be 1 pixel wide, and as such the fractional coverages are given by
\begin{subequations}
\begin{align}
    r^{\mathrm{lower}} &= \frac{\lambda^{\mathrm{upper}}-\lambda^{\mathrm{line}}}{\lambda^{\mathrm{upper}}-\lambda^{\mathrm{lower}}} \label{Eq:15a} \\
    r^{\mathrm{upper}} &= 1 - r^{\mathrm{lower}} \,,                                                                                 \label{Eq:15b}
\end{align}
\end{subequations}
where $\lambda^{\mathrm{line}}$ is the shifted central wavelength of a mask line.

The input spectra are iteratively read and interpolated onto the same wavelength grid as the first input spectrum. The CCF for the $i^{\mathrm{th}}$ spectrum, the $j^{\mathrm{th}}$ spectral order, and the $k^{\mathrm{th}}$ velocity shift is computed as the following sum over all lines $l$ in the line mask,
\begin{equation}\label{Eq:16}
    \mathrm{CCF}_{i,j}(v_{k}) = \sum_{l} w_{j,l}\left( F_{i}\left(\lambda_{j,k,l}^{\mathrm{lower}}\right)r_{j,k,l}^{\mathrm{lower}} + F_{i}\left(\lambda_{j,k,l}^{\mathrm{upper}}\right)r_{j,k,l}^{\mathrm{upper}} \right) \,,
\end{equation}
where $w$ are the weights of the spectral lines, normalized such that $\sum_{l}w_{j,l}\,{=}\,1$. By default, the weights are all equal, however, users can select any column with numbers from the line mask, suggestively, the line depth or the RV content (see Eq.~\ref{Eq:04}) as has been used for the weighted CCF in the ESPRESSO pipeline. The associated CCF uncertainties are given by
\begin{equation}\label{Eq:17}
\begin{split}
    &\sigma_{\mathrm{CCF},i,j}(v_{k}) = \\
    &\sqrt{ \sum_{l} w_{j,l}^{2} \left( \left(\sigma_{F,i}\left(\lambda_{j,k,l}^{\mathrm{lower}}\right)r_{j,k,l}^{\mathrm{lower}}\right)^{2} + \left(\sigma_{F,i}\left(\lambda_{j,k,l}^{\mathrm{upper}}\right)r_{j,k,l}^{\mathrm{upper}}\right)^{2} \right) } \,.
\end{split}
\end{equation}
Once the CCFs of the individual orders have been computed, order-summed CCFs are also produced,
\begin{equation}\label{Eq:18}
    \mathrm{CCF}_{i}(v_{k}) = \sum_{j} \mathrm{CCF}_{i,j}(v_{k}) \,,
\end{equation}
whose uncertainties are the quadratic sums of the individual CCF uncertainties,
\begin{equation}\label{Eq:19}
    \sigma_{\mathrm{CCF},i}(v_{k}) = \sqrt{ \sum_{j} \sigma_{\mathrm{CCF},i,j}(v_{k})^{2} } \,.
\end{equation}
Finally, both the individual and order-summed CCFs are fitted with inverted Gaussians (see Sect.~\ref{Sect:3.1.4}), from which the centers---corresponding to the best-fitting RVs---are extracted. The RV uncertainties,
\begin{equation}\label{Eq:20}
    \sigma_{\mathrm{RV},i,j} = \left( \sqrt{ \sum_{k} \frac{1}{\sigma_{v,i,j}(v_{k})^{2}} } \right)^{-1} \,,
\end{equation}
are propagated from the velocity uncertainties of the CCF points, $\sigma_{v}(v)$, which are estimated from the CCF gradient,
\begin{equation}\label{Eq:21}
\begin{split}
    \sigma_{v,i,j}(v_{k}) &= \left| \frac{\mathrm{d}v}{\mathrm{d}\mathrm{CCF}_{i,j}(v)} \right|_{v_{k}} \sigma_{\mathrm{CCF},i,j}(v_{k}) \\
    &\approx \left| \frac{\mathrm{d}\mathrm{CCF}_{i,j}(v)}{\mathrm{d}v} \right|^{-1}_{v_{k}} \sigma_{\mathrm{CCF},i,j}(v_{k}) \,.
\end{split}
\end{equation}
The amplitude (sometimes referred to as the contrast) and FWHM of the best-fit inverted Gaussian are also saved since they can be used as stellar activity indicators. Their uncertainties are taken as the square roots of the diagonal elements in the covariance matrix outputted by the numerical solver. In addition to the CCF, its bisector, defined as the velocity midway points at certain flux depths, are also computed and stored. The bisector is evaluated at continuum-normalized flux depths between $0$ and $0.99$ in steps of $0.01$ by interpolating the left and right sides (with respect to the minimum of the best-fit Gaussian) of the CCF and computing the average of the interpolated velocities. Flux points outside the maximal depth are assigned \acrfull*{NaN} velocities. The velocity uncertainties of the bisector are propagated from the interpolated velocity uncertainties of the CCF. Bottom and top parts of the bisector are thereafter defined as the flux intervals covering $10$--$\SI{40}{\percent}$ and $60$--$\SI{90}{\percent}$, respectively, of the bisector maximal depth, as done in \cite{Lafarga+2020}. The \acrlong*{BIS} \citep[BIS;][]{Queloz+2001}, another commonly used stellar activity proxy, is then computed as the difference between the average velocities of the bisector top and bottom parts. The BIS uncertainty is given by the square root of the quadratic sum of the top and bottom uncertainties, taken as the average bisector uncertainty within each part divided by the square root of the number of points within each part.

Our CCF computation is similar to the methodology of \cite{Lafarga+2020}. The advantage of shifting the mask lines and identifying their overlaps with neighboring pixels, is that the stellar spectrum does not have to be additionally interpolated. Although we interpolate the stellar spectra once to achieve a shared wavelength grid such that the wavelength indices do not have to be identified for each spectrum, the adopted method remains a more rigorous solution than interpolating the stellar spectra a second time onto the shifted mask line positions, and is computationally cheap in the sense that the CCF equations (Eqs.~\ref{Eq:16}--\ref{Eq:17}) has two simple arithmetic terms per mask line (for its lower and upper neighboring pixel) instead of only one.

\subsubsection{\texttt{data.compute\_vrad\_lbl()}}\label{Sect:3.2.4}

The \texttt{compute\_vrad\_lbl()} function extracts RVs from input spectra using the \acrlong*{LBL} \citep[LBL;][]{Dumusque2018,Cretignier+2020,Artigau+2022} or \acrlong*{PBP} \citep[PBP;][]{AlMoulla+2022} methods. We note that the PBP method is a generalization of the LBL method in which each spectral line is divided into parts based on the average formation temperature of each spectral point, however, we name the function after the LBL method since it is more established in the terminology. 

These methods are based upon template matching \citep{Bouchy+2001,Anglada-Escude&Butler2012,Zechmeister+2018}, in which the Doppler shift of a spectrum is derived by comparing it to a reference spectrum (also called the template). The reference in our case is a high \acrfull*{S/N} spectrum built by averaging all or a subset of the observed spectra (handled by the \texttt{compute\_spec\_reference()} function, whose extensive description is omitted in this paper due to its simplicity). Given a small velocity shift, $v$, relative to the size of the spectral features, the flux of the shifted spectrum, $\tilde{F}$, can be expressed as a Taylor expansion of the flux of the reference spectrum, $F$,
\begin{equation}\label{Eq:22}
    \tilde{F} = F - \frac{\mathrm{d}F}{\mathrm{d}\lambda}\mathrm{d}\lambda = F - \frac{\mathrm{d}F}{\mathrm{d}\lambda}\frac{v}{c}\lambda \,,
\end{equation}
where the change in wavelength, $\mathrm{d}\lambda\,{=}\,\tilde{\lambda}{-}\lambda$, is expressed in terms of the velocity using Eq.~\ref{Eq:10}. If the flux is recorded as a photo-electron count or similar unit, the shifted spectrum and reference spectrum could have different continuum levels due to differences in their observational configurations (e.g., different exposure times). Therefore, Eq.~\ref{Eq:22} is complemented with a scaling factor, $A$,
\begin{equation}\label{Eq:23}
    \tilde{F} = A \left( F - \frac{\mathrm{d}F}{\mathrm{d}\lambda}\frac{v}{c}\lambda \right) \,.
\end{equation}
Like the CCF method (see Sect.~\ref{Sect:3.2.3}), the relevant matrix indices are identified and stored in advance. Using the auxiliary data, the index ranges of each spectral line and each specified formation temperature bin are found. If no temperature bins are provided, the function uses one bin covering all formations temperatures, making the calculation equivalent to the traditional LBL method. Ultimately, it is the intersections between line and temperature indices that are used to define each spectral segment.\footnote{Note that although the same indices can be used for all spectra since they will be interpolated on a common wavelength grid, it is not warranted that the same indices can be used to the define the bounds of a line (or line part) if that line has been shifted too far away from its rest wavelength. It is therefore implied that the indices will be the same as long as the expected velocity shifts are sub-pixel, i.e., around or smaller than about $\SI{1}{\kilo\meter\per\second}$.} The template matching is then performed according to the following steps for each spectrum and each segment:
\begin{enumerate}
    \item Using the stored indices, extract the wavelength and flux points for the considered segment in the shifted spectrum.
    \item Shift the wavelengths with an initial guess on the velocity (always assumed to be $0$ the first time).
    \item Interpolate the reference spectrum and its gradient onto the same wavelength points as the segment.
    \item Compute the two unknown variables in Eq.~\ref{Eq:23}, $v$ and $A$, using a numerical least-squares solver.
    \item (Optional) Repeat steps $1$--$4$ with $v$ as the initial guess on the shift in order to converge, since the validity of the Taylor expansion decreases with the velocity amplitude.
\end{enumerate}
Assuming the reference to be essentially noiseless, the velocity uncertainty of each point, $i$, in a segment is given by
\begin{equation}\label{Eq:24}
    \sigma_{v,i} = \left| \frac{\mathrm{d}v}{\mathrm{d}\tilde{F}} \right|_{i} \sigma_{\tilde{F},i} \approx \left| \frac{\mathrm{d}\tilde{F}}{\mathrm{d}v} \right|_{i}^{-1} \sigma_{\tilde{F},i} \approx \left| \frac{\mathrm{d}F}{\mathrm{d}\lambda} \right|_{i}^{-1}\frac{c}{\lambda_{i}} \sigma_{\tilde{F},i} \,,
\end{equation}
where $\sigma_{\tilde{F}}$ are the flux uncertainties of the shifted spectrum, and where in the last step we have assumed that the flux gradient of the shifted spectrum can be approximated by the one of the reference, partially in order to not be repeatedly recomputed for each observation and partially because the high-S/N reference should yield a smoother derivative. The total RV uncertainty then becomes
\begin{equation}\label{Eq:25}
    \sigma_{\mathrm{RV}} = \left( \sqrt{ \sum_{i} \frac{1}{\sigma_{v,i}^{2}} } \right)^{-1} \,.
\end{equation}
The individual segment RVs are stored in an array with dimensions $(N_{\mathrm{ord}}){\times}(N_{\mathrm{spec}},N_{\mathrm{line}},N_{\mathrm{bin}})$, where the variables denote the number of spectral orders, spectra, spectral lines, and formation temperature bins, respectively. The cross symbol indicates that the array does not have equally long vectors along all dimensions since different spectral orders will contain distinct amounts of lines.\footnote{This is the only instance where $N_{\mathrm{ord}}$ precedes $N_{\mathrm{spec}}$ in the dimensionality order of a time serial array. The array is structured this way to enable it to be concatenable across all spectral orders if the user desires a LBL array with normal indexation.} In addition, the function also computes a weighted average of all lines stored in an array with dimensions $(N_{\mathrm{spec}},N_{\mathrm{ord}},N_{\mathrm{bin}})$, a weighted average of all lines and orders in an array with dimensions $(N_{\mathrm{spec}},N_{\mathrm{bin}})$, and a default time series with dimension $N_{\mathrm{spec}}$ taken to be the first bin of the line- and order-averaged RV array.

\subsection{The \texttt{star} subclass}\label{Sect:3.3}

The \texttt{star} subclass handles the storage of stellar parameters, the computation of the \acrfull*{VPSD}, and the characterization of stellar activity signals in the VPSD.

The \texttt{star} class variables are the following:
\begin{labeling}[~:]{\texttt{stellar\_parameters}}
	\item[\texttt{target}             ] the star name
    \item[\texttt{stellar\_parameters}] the stellar parameters
    \item[\texttt{vpsd}               ] the VPSD
    \item[\texttt{vpsd\_components}   ] the VPSD components
\end{labeling}

\subsubsection{\texttt{star.get\_stellar\_parameters()}}\label{Sect:3.3.1}

The \texttt{get\_stellar\_parameters()} function retrieves approximate stellar parameters for the target star. If the \texttt{target} variable is specified, the function queries the spectral type and systematic RV from SIMBAD\footnote{SIMBAD stands for \acrlong*{SIMBAD}.\\Available at \url{https://simbad.u-strasbg.fr/simbad}}. Based on the queried spectral type, the stellar parameters are approximated by interpolating the values in Table~\ref{Tab:01} and thereafter computing the micro- and macroturbulences from Eqs.~\ref{Eq:01}--\ref{Eq:02}.

\subsubsection{\texttt{star.compute\_vpsd()}}\label{Sect:3.3.2}

The \texttt{compute\_vpsd()} function computes the VPSD from the provided or extracted RVs by computing an unnormalized GLS periodogram using \texttt{functions.gls\_periodogram()}. To convert the power spectrum into a density, i.e., a quantity independent of measurement sampling and duration, the outputted periodogram amplitude is divided by the area of the window function. The window function is obtained from the GLS periodogram of a single, unity-amplitude sinusoid, with a frequency equal to the center-most sampled frequency by the periodogram, evaluated at the same time stamps as the measured signal. Additionally, the VPSD is averaged over a user-defined number (defaults to $50$) of logarithmically equidistant frequency bins.

\subsubsection{\texttt{star.add\_vpsd\_components()}}\label{Sect:3.3.3}

\begin{figure}[ht!]
    \includegraphics[width=\linewidth]{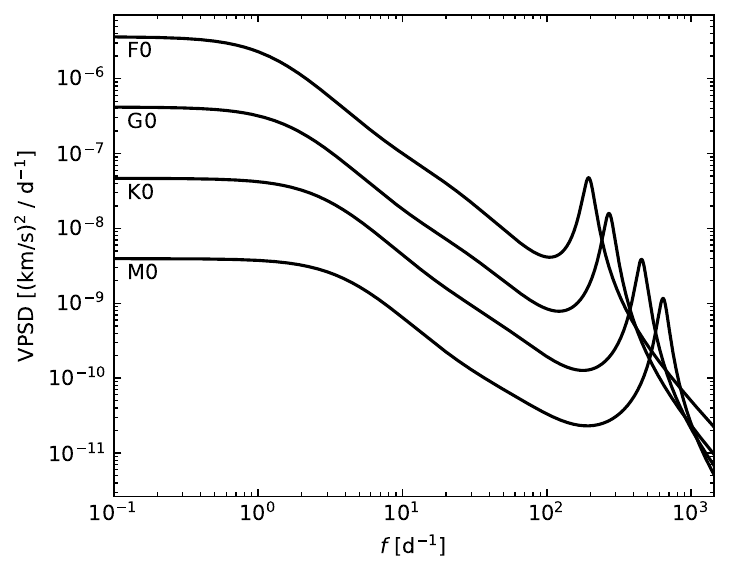}
    \caption{Example of VPSDs for spectral types F0, G0, K0, and M0. The sampled frequencies correspond to an RV time series with a span of $\SI{10}{\day}$ and a median spacing of $\SI{30}{\second}$, hence the VPSDs resolve the oscillations and both granulation phenomena for all shown spectral types.}
    \label{Fig:02}
\end{figure}

The \texttt{add\_vpsd\_components()} function manually or automatically determines which stellar signal components are present in the VPSD. Either the user provides a list of the component names to be included, or the function assumes all components are present and adds them if their characteristic frequencies are within the frequency bounds of the VPSD. The considered components are oscillations, granulation, and supergranulation. A photon noise term is also included to capture all features in the VPSD. The components are stored as the coefficients of analytical functions. We follow the formalism of \cite{Lefebvre+2008}, where they modeled the envelope of oscillation modes with a Lorentzian function, each granulation phenomenon with a Harvey function, and the photon noise with a frequency-independent offset,
\begin{equation}\label{Eq:26}
\begin{split}
    &\mathrm{VPSD}(f) = \mathrm{VPSD}_{\mathrm{osc}} + \sum_{i} \mathrm{VPSD}_{\mathrm{gra},i} + \mathrm{VPSD}_{\mathrm{noise}}\\
    &= A_{\mathrm{osc}}\frac{\gamma^{2}}{\gamma^{2} + (f-f_{\mathrm{max}})^{2}} + \sum_{i} A_{\mathrm{gra},i}\frac{1}{1+(\tau_{i}f)^{\alpha_{i}}} + A_{\mathrm{noise}} \,,
\end{split}
\end{equation}
where the indices $i\,{=}\,1,2$ represent granulation and supergranulation, respectively. Here, $A_{\mathrm{osc}}$, $A_{\mathrm{gra}}$, and $A_{\mathrm{noise}}$ are the oscillation, (super)granulation, and photon noise amplitudes, respectively, $\gamma$ and $f_{\mathrm{max}}$ are the oscillation envelope \acrfull*{HWHM} and central frequency, and $\tau$ and $\alpha$ are the (super)granulation characteristic timescales and decay rates. \cite{Guo+2022} used analytical and empirical photometric relations from the literature \citep[primarily][]{Kjeldsen&Bedding1995,Kjeldsen&Bedding2011,Kallinger+2014,Basu&Chaplin2017} in order to describe all the coefficients (except the decay rates) as a function of $f_{\mathrm{max}}$, which itself only depends on the stellar parameters,
\begin{equation}\label{Eq:27}
    f_{\mathrm{max}} = f_{\mathrm{max},\Sun}\left(\frac{T_{\mathrm{eff}}}{T_{\mathrm{eff},\Sun}}\right)^{-1/2}\frac{g}{g_{\Sun}} \,,
\end{equation}
where variables with subscript $\Sun$ are the solar counterparts. \cite{Guo+2022} were also able to convert between photometry and velocimetry using the following scaling relations\footnote{Note that these scaling relations are technically optimized for bolometrically corrected photometry and optical velocimetry, and might slightly deviate for different bandpasses or wavelength ranges.} for the oscillation and granulation amplitudes,
\begin{subequations}
\begin{align}
    &\frac{A_{\mathrm{osc},\mathrm{velocimetry}}}{A_{\mathrm{osc},\mathrm{photometry}}} \, \propto \left(\frac{T_{\mathrm{eff}}}{T_{\mathrm{eff},\Sun}}\right)^{1.8} \equiv r_{\mathrm{osc}} \,,                                        \label{Eq:28a} \\
    &\frac{A_{\mathrm{gra},\mathrm{velocimetry}}}{A_{\mathrm{gra},\mathrm{photometry}}} \, \propto \left(\frac{T_{\mathrm{eff}}}{T_{\mathrm{eff},\Sun}}\right)^{64/9}\left(\frac{g}{g_{\Sun}}\right)^{-4/9} \equiv r_{\mathrm{gra}} \,. \label{Eq:28b}
\end{align}
\end{subequations}
Although \cite{Guo+2022} provides a way of computing the value of each coefficient using only $f_{\mathrm{max}}$, we found that some coefficients were poorly translated to velocimetric measurements of the Sun. For example, the characteristic timescale of granulation is estimated to be $\tau_{1}\,{=}\,(f_{\mathrm{max},\Sun}/\SI{}{\micro\hertz})^{-0.970}/0.317\,\SI{}{\per\micro\hertz}\,{\approx}\,\SI{20}{\minute}$, which is considerably shorter than the typical timescale of about $\SI{1}{\hour}$ found by RV surveys of the Sun and Sun-like stars \citep{Dumusque+2011,AlMoulla+2023}. Therefore, we decided to use the scaling relations from \cite{Guo+2022} to extend to other spectral types, but to calibrate the coefficients using the solar values from \cite{AlMoulla+2023}, converted to the units used by \texttt{ARVE} and summarized in Table~\ref{Tab:02}. For the oscillation coefficients, we derive the following scaling relations for the amplitude and HWHM,
\begin{equation}\label{Eq:29}
    A_{\mathrm{osc}} = A_{\mathrm{osc},\Sun} r_{\mathrm{osc}} \left(\frac{f_{\mathrm{max}}}{f_{\mathrm{max},\Sun}}\right)^{-2.305} \,,
\end{equation}
\begin{equation}\label{Eq:30}
    \gamma = \gamma_{\Sun} \left(\frac{f_{\mathrm{max}}}{f_{\mathrm{max},\Sun}}\right)^{0.880} \,.
\end{equation}
For the (super)granulation coefficients, we derive the following scaling relations for the amplitude and characteristic timescale,
\begin{subequations}
\begin{align}
    A_{\mathrm{gra},1} &= A_{\mathrm{gra},1,\Sun} r_{\mathrm{gra}} \left(\frac{f_{\mathrm{max}}}{f_{\mathrm{max},\Sun}}\right)^{-2.188} \,, \label{Eq:31a} \\
    A_{\mathrm{gra},2} &= A_{\mathrm{gra},2,\Sun} r_{\mathrm{gra}} \left(\frac{f_{\mathrm{max}}}{f_{\mathrm{max},\Sun}}\right)^{-2.210} \,, \label{Eq:31b}
\end{align}
\end{subequations}
\begin{subequations}
\begin{align}
    \tau_{1} &= \tau_{1,\Sun} \left(\frac{f_{\mathrm{max}}}{f_{\mathrm{max},\Sun}}\right)^{-0.970} \,, \label{Eq:32a} \\
    \tau_{2} &= \tau_{2,\Sun} \left(\frac{f_{\mathrm{max}}}{f_{\mathrm{max},\Sun}}\right)^{-0.992} \,. \label{Eq:32b}
\end{align}
\end{subequations}
The decay rates are fixed to $2$ as in \cite{AlMoulla+2023}. For the photon noise amplitude, it is simply taken to be the minimum value of the VPSD of the input RVs. Finally, solely components whose characteristic frequency is within the sampled range are added, i.e., oscillations or (super)granulation are only considered if $f_{\mathrm{max}}$ or $1/\tau$ are within the frequency bounds of Eq.~\ref{Eq:13}. An example of how the total VPSD (without any photon noise) looks like for different spectral types is shown in Fig.~\ref{Fig:02}.

\begin{table*}[t!]
	\caption{Best-fit solar VPSD coefficients from \cite{AlMoulla+2023}, converted to the units used by \texttt{ARVE}.}
	\begin{tabular*}{\textwidth}{l @{\extracolsep{\fill}} l @{\extracolsep{\fill}} l @{\extracolsep{\fill}} l}
		\toprule
		\midrule
		\textbf{Component}                & \textbf{Coefficient}      & \textbf{Value}   & \textbf{Unit}                                           \\
		\midrule
		\multirow{3}{*}{Oscillations}     & $A_{\mathrm{osc},\Sun}$   & $\SI{1.45e-8}{}$ & $\SI{}{\kilo\meter\squared\per\second\squared\per\day}$ \\
		& $\gamma_{\Sun}$           & $\SI{2.50e1 }{}$ & $\SI{}{\per\day}$                                       \\
		& $f_{\mathrm{max},\Sun}$   & $\SI{2.73e2 }{}$ & $\SI{}{\per\day}$                                       \\
		\midrule
		\multirow{2}{*}{Granulation}      & $A_{\mathrm{gra},1,\Sun}$ & $\SI{4.76e-9}{}$ & $\SI{}{\kilo\meter\squared\per\second\squared\per\day}$ \\
		& $\tau_{1,\Sun}$           & $\SI{3.80e-2}{}$ & $\SI{}{\day}$                                           \\
		\midrule
		\multirow{2}{*}{Supergranulation} & $A_{\mathrm{gra},2,\Sun}$ & $\SI{3.23e-7}{}$ & $\SI{}{\kilo\meter\squared\per\second\squared\per\day}$ \\
		& $\tau_{2,\Sun}$           & $\SI{5.54e-1}{}$ & $\SI{}{\day}$                                           \\
		\bottomrule
	\end{tabular*}
	\label{Tab:02}
\end{table*}

\subsubsection{\texttt{star.fit\_vpsd\_components()}}\label{Sect:3.3.4}

The \texttt{fit\_vpsd\_components()} function computes the best-fitting VPSD component coefficients in Eq.~\ref{Eq:26} through a Levenberg–Marquardt minimization of the absolute residuals between the logarithm of Eq.~\ref{Eq:26} (the sum of all the analytical VPSD components) and the logarithm of the VPSD of the input RVs. The motivation for using logarithmic instead of linear residuals is because the VPSD typically spans several orders of magnitude, and therefore we do not want relatively small improvements of the high-amplitude granulation terms to be favored over relatively large improvements of the low-amplitude oscillation and noise terms.

\subsubsection{\texttt{star.simulate\_vrad\_from\_vpsd\_components()}}\label{Sect:3.3.5}

The \texttt{simulate\_vrad\_from\_vpsd\_components()} function enables the simulation of RVs of individual or combined VPSD components on an evenly sampled, user-defined time grid. The RV of any given activity component is given by discretely integrating its VPSD analytical function evaluated at the frequencies given by Eq.~\ref{Eq:13}, multiplied with sinusoids of the same frequencies with randomized phases $\phi$,
\begin{equation}\label{Eq:33}
    \mathrm{RV}_{\mathrm{comp}}(t) = \sum_{i} \sqrt{\mathrm{VPSD}_{\mathrm{comp}}(f_{i})\mathrm{d}f}\sin(2{\pi}f_{i}t + \phi_{i}) \,,
\end{equation}
where $\mathrm{VPSD}_{\mathrm{comp}}$ is the VPSD of the considered component, and $\mathrm{d}f$ is the frequency step. Calling the function simulates RVs for each component added by the \texttt{star.add\_vpsd\_components()} function, as well as RVs for their combined VPSDs, both with and without photon noise. The ability to simulate activity components from the frequency to the temporal domain is useful to determine, e.g., the \acrfull*{RMS} contribution of each component to the total RV dispersion, or to have an easily modifiable RV time series on which activity mitigation and observational strategies can be trialed and optimized \citep[e.g.,][]{Dumusque+2011,Luhn+2023}.

\subsection{The \texttt{planets} subclass}\label{Sect:3.4}

The \texttt{planets} subclass deals with Keplerian signals, which could be interpreted to be of planetary nature, detected from or injected into the RV time series.

The \texttt{planets} class variables are the following:
\begin{labeling}[~:]{\texttt{periodograms}}
	\item[\texttt{periodograms}] the GLS periodograms
    \item[\texttt{keplerians}  ] the fitted Keplerians
    \item[\texttt{recoveries}  ] the recovered injected planets
\end{labeling}

\subsubsection{\texttt{planets.fit\_keplerians()}}\label{Sect:3.4.1}

The \texttt{fit\_keplerians()} function computes a normalized GLS periodogram (see Sect.~\ref{Sect:3.1.5}) of the stored RV time series, finds the largest periodogram peak, fits a Keplerian signal with that period in the RVs, stores its parameters, and thereafter subtracts the Keplerian from a copy of the RVs. These steps are repeated until there are no more peaks above a user-defined \acrfull*{FAP} level (defaults to $\SI{1}{\percent}$) or until the number of fitted Keplerians reaches a user-defined maximum (defaults to $10$). The fitted Keplerians are assumed to be circular of the form,
\begin{equation}\label{Eq:34}
    \mathrm{RV}(t) = K\sin\left(\frac{2{\pi}t}{P} + \phi\right) + C
\end{equation}
where $K$ is the RV semi-amplitude, $P$ the period, $\phi$ the phase, and $C$ a constant to allow for an off-set in case the RVs are not centered around zero. Future versions of \texttt{ARVE} might incorporate non-circular orbits, however, the current priorities of the code is to provide the user with a fast tool to investigate planetary detection limits (see Sect.~\ref{Sect:4.2}), rather than robustly fitting any type of Keplerian. If the latter is needed, the reader is referred to other existing Python packages, such as \texttt{RadVel} \citep{Fulton+2018}.

\subsubsection{\texttt{planets.injection\_recovery()}}\label{Sect:3.4.2}

The \texttt{injection\_recovery()} function injects planetary signals according to Eq.~\ref{Eq:34} (with a zero off-set) into a copy of the stored RVs and attempts to recover them. The user can either specify an array or a grid of planetary parameters. If an array is selected, the planets are all injected at once, which can be interpreted as a planetary system. If a grid is selected, the planets are injected one by one, to explore the parameter space. For the planetary parameters, the user can specify either the periods, $P$, or equivalently the semi-major axes, $a$, related by
\begin{equation}\label{Eq:35}
\begin{split}
    &P = \frac{2{\pi}a^{3/2}}{(G(M+m))^{1/2}} \approx \frac{2{\pi}a^{3/2}}{(GM)^{1/2}}\\
    &\approx \SI{365.25}{} \, \left(\frac{a}{\SI{}{\au}}\right)^{3/2}\left(\frac{M}{M_{\Sun}}\right)^{-1/2} \, \SI{}{\day} \,,
\end{split}
\end{equation}
where $G$ is the gravitational constant and $a$ is measured in \acrfullpl*{AU}; the user can also specify either the RV semi-amplitudes, $K$, or equivalently the masses, $m$, related by
\begin{equation}\label{Eq:36}
\begin{split}
    &K = \left(\frac{2{\pi}G}{P}\right)^{1/3}\frac{m}{(M+m)^{2/3}}\frac{\sin{i}}{(1-e^{2})^{1/2}}\\
    &\approx \left(\frac{2{\pi}G}{P}\right)^{1/3}\frac{m}{M^{2/3}}\frac{\sin{i}}{(1-e^{2})^{1/2}}\\
    &\approx \SI{8.95}{} \, \left(\frac{P}{\SI{365.25}{\day}}\right)^{-1/3}\left(\frac{M}{M_{\Sun}}\right)^{-2/3}\frac{m}{m_{\Earth}}\frac{\sin{i}}{(1-e^{2})^{1/2}} \, \SI{}{\centi\meter\per\second} \,,
\end{split}
\end{equation}
where $i$ and $e$ are the planets' orbital inclinations and eccentricities, respectively, and subscript $\Earth$ denotes properties of the Earth. In case of an array of planets, their phases can also be specified; if not, they will be randomized, as in the case of a grid. Note that the injected masses (or RV semi-amplitudes) are by default minimum masses, $m\sin{i}$, because no assumption on the inclination is made, and since \texttt{ARVE} currently only handles circular orbits, the eccentricity is always zero.

The recovery process is performed by calling the \texttt{planets.fit\_keplerians()} function. The injected planet (or planets if multiple are injected simultaneously) is defined to be recovered if any of the fitted Keplerian periods are within a user-defined interval around the true period (defaults to within $\SI{10}{\percent}$ of the true period); if several Keplerians satisfy this criterion, the one with the closest RV semi-amplitude is chosen. For each injected planet, the function returns the ratio between the recovered and injected RV semi-amplitudes if the planet is recovered, and NaN otherwise.

\section{Demonstration on HARPS-N and NEID solar data}\label{Sect:4}

To showcase some of the capabilities of \texttt{ARVE}, we provide two demonstrations on how the code can be used to retrieve valuable characterization of RV time series. The examples make use of data from the HARPS-N solar telescope \citep{Dumusque+2015,Phillips+2016} and the NEID solar telescope \citep{Lin+2022}. The HARPS-N solar data was observed from July 2015 to July 2018, and consists of S1D spectra, RVs and activity indicators delivered by the official \acrlong*{DRS} \citep[DRS;][]{Dumusque+2021} version \texttt{3.0.1} and curated by a Bayesian mixture model to filter out measurements taken during poor weather conditions when the resolved Sun was partially covered by clouds \citep{CollierCameron+2019}. The data products are downloadable from DACE\footnote{DACE stands for \acrlong*{DACE}.\\Available at \url{https://dace.unige.ch/sun}}. The NEID solar data was observed from January 2021 to June 2024, and consists of RVs and activity indicators delivered by the official \acrfull*{DRP}\footnote{Available at\\\url{https://neid.ipac.caltech.edu/docs/NEID-DRP/}} version \texttt{1.3} and filtered as described in \cite{Ford+2024}. The filtered data products are downloadable from Zenodo\footnote{Available at\\\url{https://zenodo.org/records/13363762}} and the full data set is downloadable from the NEID Solar RV Archive\footnote{Available at\\\url{https://neid.ipac.caltech.edu/search_solar.php}}.

\subsection{Evolution of granulation timescales}\label{Sect:4.1}

The solar magnetic cycle, with an average length of $11$ years, modulate the number and preferred latitudes of magnetically active surface regions, such as bright faculae and dark spots. However, whether magnetic cycles, in the Sun or other stars, also affect the properties of higher-frequency stellar variability, such as oscillations and granulation phenomena, have been poorly studied.

\begin{figure}[ht!]
    \includegraphics[width=\linewidth]{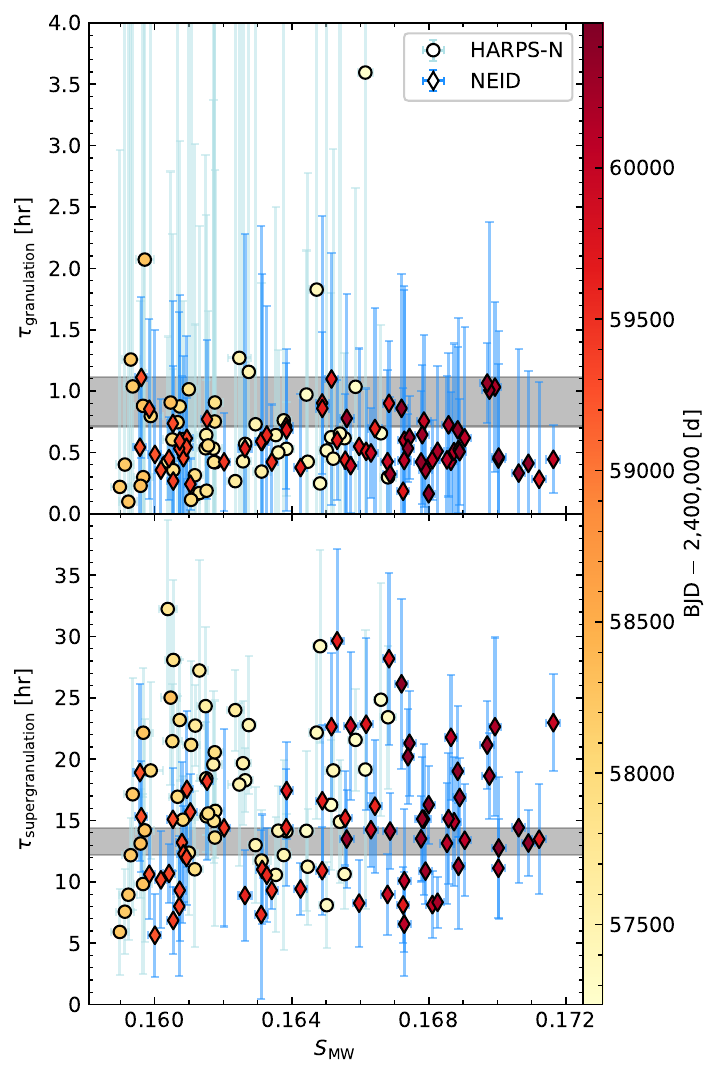}
    \caption{Solar granulation characteristic timescales at different activity levels. \textit{Upper panel:} The best-fit granulation characteristic timescale computed on $\SI{10}{\day}$-intervals of the HARPS-N (circles with light blue error bars) and NEID (diamonds with dark blue error bars) solar RV time series. The abscissa shows the activity level as traced by the average $S_{\mathrm{MW}}$ index within each interval. The markers are color-coded by the average time of the points within each interval. The gray-shaded region shows the best-fit granulation timescale with ${\pm}1\sigma$ uncertainties obtained by \cite{AlMoulla+2023} from a combined data set of HARPS-N and HARPS solar RVs. \textit{Lower panel}: Same as the upper panel, but for the supergranulation timescale.}
    \label{Fig:03}
\end{figure}

We make use of \texttt{ARVE}'s ability to compute the VPSD of an RV time series, fit it with analytical functions for the relevant activity components, and extract the values for the coefficients of interest from said analytical expressions. We apply these steps to the HARPS-N and NEID RVs individually. The VPSD coefficients are fitted on separate $\SI{10}{\day}$-intervals, from which we extract the characteristic timescales of the granulation and supergranulation components in order to evaluate how they change over time and at various activity levels. The results are shown in Fig.~\ref{Fig:03}, where the timescales are plotted against the average Mount Wilson $S_{\mathrm{MW}}$ index of each interval. The granulation timescale shows no correlation with activity, whereas the supergranulation timescale seems to be prolonged with heightened activity levels, albeit with a large dispersion and a weak Pearson correlation of $0.32$ for the two combined data sets. We also remark that the uncertainties of the granulation timescales are relatively large, especially for the HARPS-N data. This is due to the granulation component being poorly constrained in the absence of an oscillation envelope in the VPSD when the exposure time of the sampled RVs cancel out the oscillations, which is the case for the HARPS-N solar telescope having an exposure time of about $\SI{5}{\minute}$, but not for the NEID solar telescope having an exposure time of about $\SI{1}{\minute}$.

\begin{figure*}[ht!]
    \includegraphics[width=0.50\textwidth]{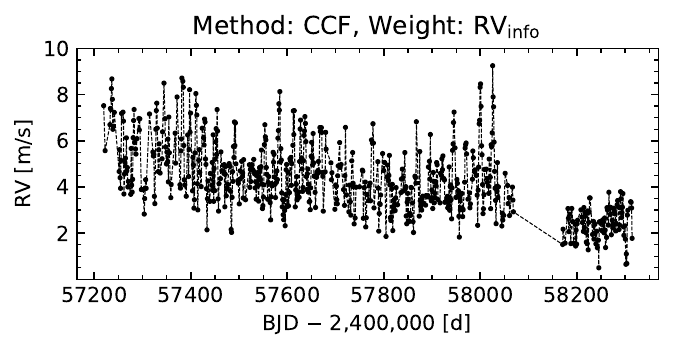}
    \includegraphics[width=0.50\textwidth]{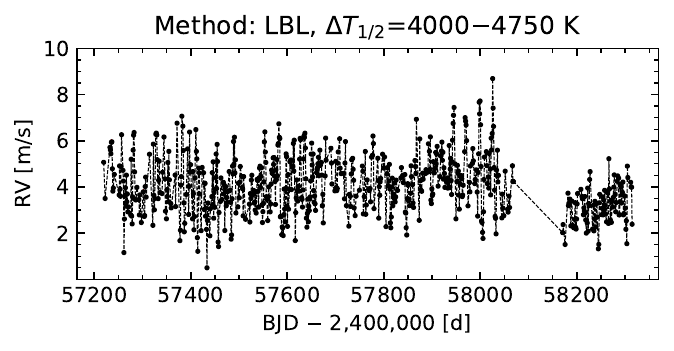}
    \includegraphics[width=0.50\textwidth]{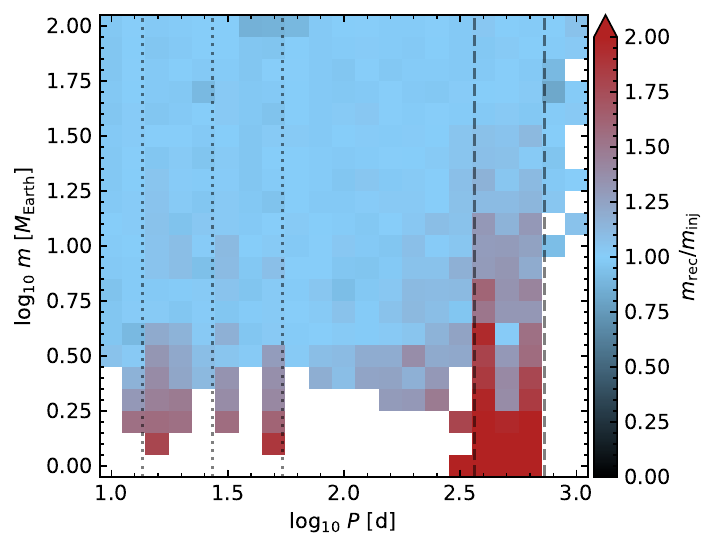}
    \includegraphics[width=0.50\textwidth]{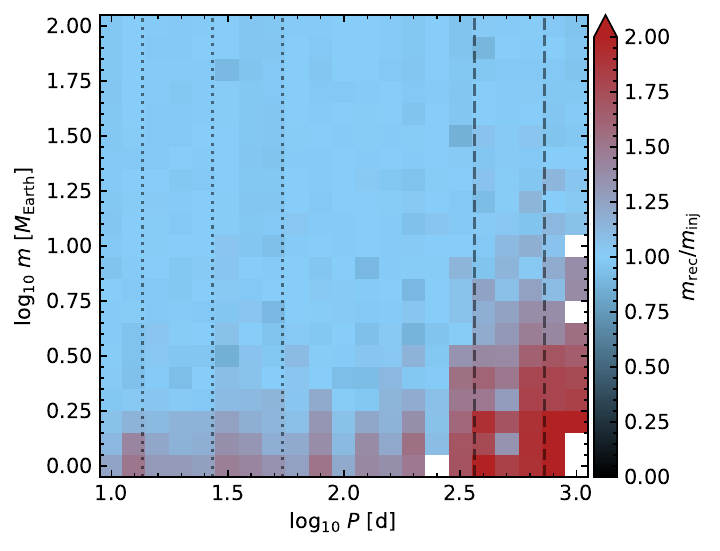}
    \includegraphics[width=0.50\textwidth]{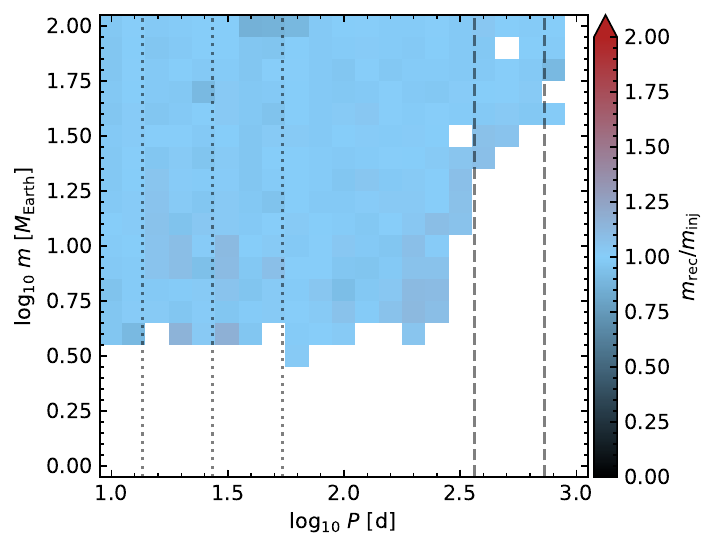}
    \includegraphics[width=0.50\textwidth]{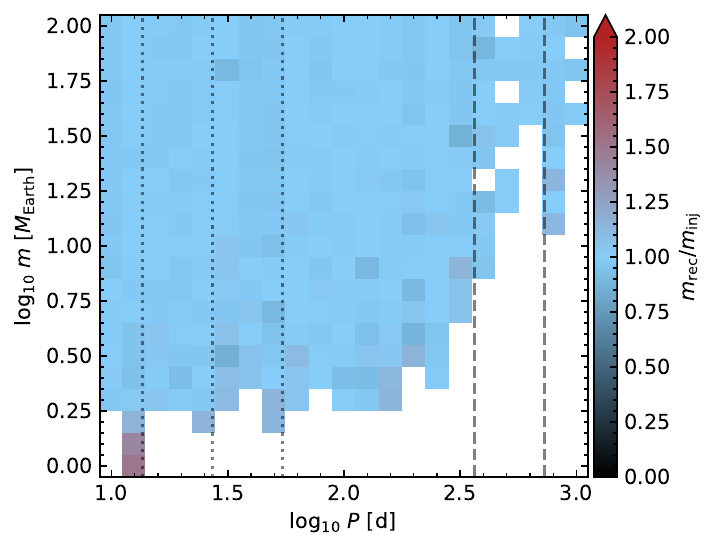}
    \caption{Planetary detection limits for two different RV extraction methods applied on the same HARPS-N data set. \textit{Upper left panel}: RVs extracted using a CCF weighted with the RV content of each mask line. \textit{Upper right panel}: RVs extracted using the LBL technique considering line parts with average formation temperatures between $\SI{4000}{}$--$\SI{4750}{\kelvin}$ (see main text for details about the choice of temperature bin). \textit{Central left panel}: Injection-recovery test applied to the CCF RVs (upper left panel) using planets with periods between $\SI{10}{}$--$\SI{1000}{\day}$ and masses between $\SI{1}{}$--$\SI{100}{}\,m_{\Earth}$. The color indicates the ratio between the recovered and injected planetary masses for planets considered to be recovered; parameter combinations in white indicate planets which were not recovered. The displayed results are the average of $10$ trials with randomized orbital phases, for which the injected planets were detected in at least one of the trials. The dotted lines indicate the synodic solar rotation period, $P_{\mathrm{rot},\Sun}\,{\approx}\,\SI{27}{\day}$, its first harmonic at $P_{\mathrm{rot},\Sun}/2$ and its first multiple at $2P_{\mathrm{rot},\Sun}$, and the dashed lines indicate the Earth orbital period, $P_{\mathrm{orb},\Earth}\,{\approx}\,\SI{365}{\day}$, and its first multiple at $2P_{\mathrm{orb},\Earth}$. \textit{Central right panel}: Same as the central left panel, but applied to the LBL RVs (upper right panel). \textit{Lower left panel}: Same as the central left panel, but where the results are averaged over the injected planets which were detected in strictly all of the trials. \textit{Lower right panel}: Same as the lower left panel, but applied to the LBL RVs (upper right panel).}
    \label{Fig:04}
\end{figure*}

\subsection{Planetary detection limits for varied RV extraction}\label{Sect:4.2}

Extracting RVs with various techniques, such as the CCF and LBL techniques available with \texttt{ARVE}, not only enables the characterization of the impact of stellar activity and telluric contamination on stellar spectra, but also allows the outputted RV time series to be optimized for planetary detections. Here we make use of the HARPS-N solar spectra, which we shift to the heliocentric restframe in order to remove the influence of Solar System planets, bin daily to mitigate some of the activity and to make the following investigations computationally less expensive, and apply post-processing with \texttt{YARARA} which among other aspects corrects for instrumental systematics in the data. Note that \texttt{YARARA} also partially corrects for stellar activity, however, we reintroduce the activity correction in order to isolate its effect in relation to RV extraction.

We make use of \texttt{ARVE}'s ability to extract RV time series with different techniques, and thereafter apply an injection-recovery test on the RVs to estimate the detectable planetary parameters given the extraction method. We try two different methods: the CCF weighted with the RV content (see Eq.~\ref{Eq:04}) of each mask line, and the LBL technique where we average together the RVs of individual line parts formed between an average formation temperature of $\SI{4000}{}$--$\SI{4750}{\kelvin}$. This specific formation temperature bin is chosen because it was found to yield particularly low RV RMS values when exploring the entire line-forming temperature range \citep{Rescigno&AlMoulla2025}.

Fig.~\ref{Fig:04} shows the resulting RV time series, and the recovered-to-injected planetary mass ratios for a range of different periods and masses when fitting the periodograms with up to three peaks. We repeated the injection-recovery test $10$ times---with randomized orbital phases between trials---and used two different averaging methods to interpret the results. First, we averaged results from planets detected in at least one trial (central panels of Fig.~\ref{Fig:04}), which reveals a tendency to overestimate planetary masses near the detection limit; this happens when the injected planet's orbital period and phase coincidentally aligns with stellar activity signals like rotation or the magnetic cycle. Second, we averaged results from planets detected in strictly all trials (lower panels of Fig.~\ref{Fig:04}), which demonstrates that LBL RVs derived from lines formed at certain formation temperatures systematically recover more planets than CCF RVs across nearly all periods. We remark that simply changing the method used to extract the RVs, without any additional activity correction, can improve the capability to detect planets of certain properties. In particular, the LBL RVs for our selected temperature bin push the recovery limit toward lower masses at shorter periods and are seemingly less affected by the magnetic cycle. The LBL RVs appear to also give less overestimated planetary masses around the solar rotation period, the Earth orbital period, and their harmonics and multiples; the $1$-year signal shares periodicity with potential residual telluric contamination or seasonal instrumental systematics.

It should be remarked that although some of the recovered planetary signals in the LBL-case are seemingly outperforming current real-life detection limits---primarily due to the excellent S/N and sampling of the solar data, as well as the planets being injected at the RV level post-correction, as opposed to the spectral level pre-correction---the ability to swiftly compare detection estimates for, e.g., different RV extraction methods serves an important purpose. In a more realistic scenario, sampling gaps would make the planets near the recovery limit harder to recover, and the fitted periodogram peaks of non-planetary nature would have to be accounted for and explained (usually by instrumental and/or stellar signals and their aliases) before a planet can be claimed as detected.

\section{Discussion and conclusion}\label{Sect:5}

\subsection{A multi-functional tool for EPRV analysis}\label{Sect:5.1}

With \texttt{ARVE}, we introduce a novel and hopefully user-friendly tool with which RV extraction and analysis can be applied to a wide range of optical and NIR high-resolution spectra. \texttt{ARVE}'s pre-computed auxiliary data allows the RV extraction functions to solely require the spectra as input, and also conveniently provides physical parameters of the star and its spectral features, allowing a facilitated mean of studying the velocimetric behavior of individual lines or subsets with shared properties. We foresee that these kind of tools will be important in the coming years as EPRV analysis becomes more targeted toward the underlying physical processes of stellar variability \citep{Crass+2021}.

So far, \texttt{ARVE} has already been used in several recent publications, including an investigation into how GP hyperparameters evolve when modeled on RVs measured at various average formation temperatures \citep{Rescigno&AlMoulla2025}, a demonstration of how template-based RV extraction can introduce significant trends when the template is constructed from observations taken during a short timescale \citep{Silva+2025}, and a granulation study of the Maunder minimum star HD\,166620 \citep{AnnaJohn+2025}. Other applications could include RV extraction with tailored line masks or line weights, homogeneously analyzing stellar activity properties on a large sample of stars, or---more conceptually---exploring optimal methodological practices, such as the simulated impact of spectral sampling and interpolation on RV precision and planetary mass estimation. All this could be explored while keeping the required input as minimal as possible, i.e., ideally no more than time series of reduced spectra or RVs, and keeping the output as interpretable as possible, neatly stored in the same class variable throughout the analysis.

\texttt{ARVE} could also act as an introductory tool for new members of the EPRV community, seeking to familiarize with the concepts by exploring more than just the end-product RVs for themselves. A tool which relatively easily generates, e.g., CCFs and bisectors from simple input data could be valuable to enable the user the allocate their time on developing novel activity-mitigating approaches.

\subsection{Planned and potential improvements and additions}\label{Sect:5.2}

\texttt{ARVE}'s current functionalities are currently limited to those described in this paper. However, as the code is built in a modularized fashion, we anticipate that added functionality would require little to no modification of the current code structure and that the code can be continuously updated with either improved or new functions. These include, for example, the derivation of classical and/or new stellar activity indicators from the input spectra themselves, and the regression of these indicators onto the RVs, either with simple linear models or more sophisticated ones such as the increasingly popular GPs. These activity indicators could also be used in unison with the \texttt{planets} subclass, to further classify whether the fitted sinusoids are likely of planetary or stellar nature.

Furthermore, to make \texttt{ARVE} compatible with as many spectrographs as possible, rather than having the user manually load or save the input spectra in a specific format, the \texttt{data} subclass can be modified to read the output FITS formats of additional spectrograph reduction pipelines. Currently, \texttt{ARVE} can read FITS files from the ESPRESSO, EXPRES, HARPS, HARPS-N, NEID, NIRPS, and SPIRou spectrographs; other instruments can be added upon request.

\text{}\\
\footnotesize
\noindent
\textit{Data availability.} \texttt{ARVE} is an open-source software. It is downloadable from GitHub (\url{https://github.com/almoulla/arve}), installable through PyPI (\url{https://pypi.org/project/arve}), and documented on Read the Docs (\url{https://arve.readthedocs.io}). This paper refers specifically to version \texttt{1.0.0} of the code.

\text{}\\
\footnotesize
\noindent
\textit{Acknowledgments.} KA would like to thank the referee, Jinglin Zhao, for their valuable comments which improved the quality of the manuscript. KA also thanks Xavier Dumusque and Michael Cretignier for advice on the radial velocity algorithms, Jim Lundin for fruitful discussions about the code structure, and Mona Al Moulla for help with the \texttt{ARVE} logotype design. KA acknowledges support from the Swiss National Science Foundation (SNSF) under the Postdoc Mobility grant P500PT\_230225. This work has been carried out within the framework of the National Centre of Competence in Research (NCCR) PlanetS supported by the SNSF under grants 51NF40\_182901 and 51NF40\_205606. This project has received funding from the European Research Council (ERC) under the European Union’s Horizon 2020 research and innovation program (grant agreement SCORE No. 851555). This publication makes use of The Data \& Analysis Center for Exoplanets (DACE), which is a facility based at the University of Geneva (CH) dedicated to extrasolar planets data visualisation, exchange and analysis. DACE is a platform of the NCCR PlanetS, federating the Swiss expertise in exoplanet research. The DACE platform is available at \url{https://dace.unige.ch}. This research has made use of the SIMBAD database, operated at CDS, Strasbourg, France. This work has made use of the VALD database, operated at Uppsala University, the Institute of Astronomy RAS in Moscow, and the University of Vienna.

\texttt{ARVE} requires the following Python packages:
\texttt{astropy} \citep{AstropyCollaboration+2013}, \texttt{astroquery} \citep{Ginsburg+2021}, \texttt{lmfit} \citep{Newville+2023}, \texttt{matplotlib} \citep{Caswell+2021}, \texttt{numpy} \citep{Harris+2020}, \texttt{pandas} \citep{Pandas+2022}, \texttt{scipy} \citep{Gommers+2022}, \texttt{tqdm} \citep{daCosta-Luis+2022}. \texttt{ARVE} makes use of, but is not dependent on, the following additional Python packages: \texttt{HAPI} \citep{Kochanov+2016}, \texttt{PySME} \citep{Wehrhahn+2023}.

\balance

\bibliographystyle{aa}
\bibliography{Bibliography}

\onecolumn
\begin{appendix}

\section{Auxiliary data}\label{Sect:A}

\begin{figure*}[ht!]
    \includegraphics[width=\textwidth]{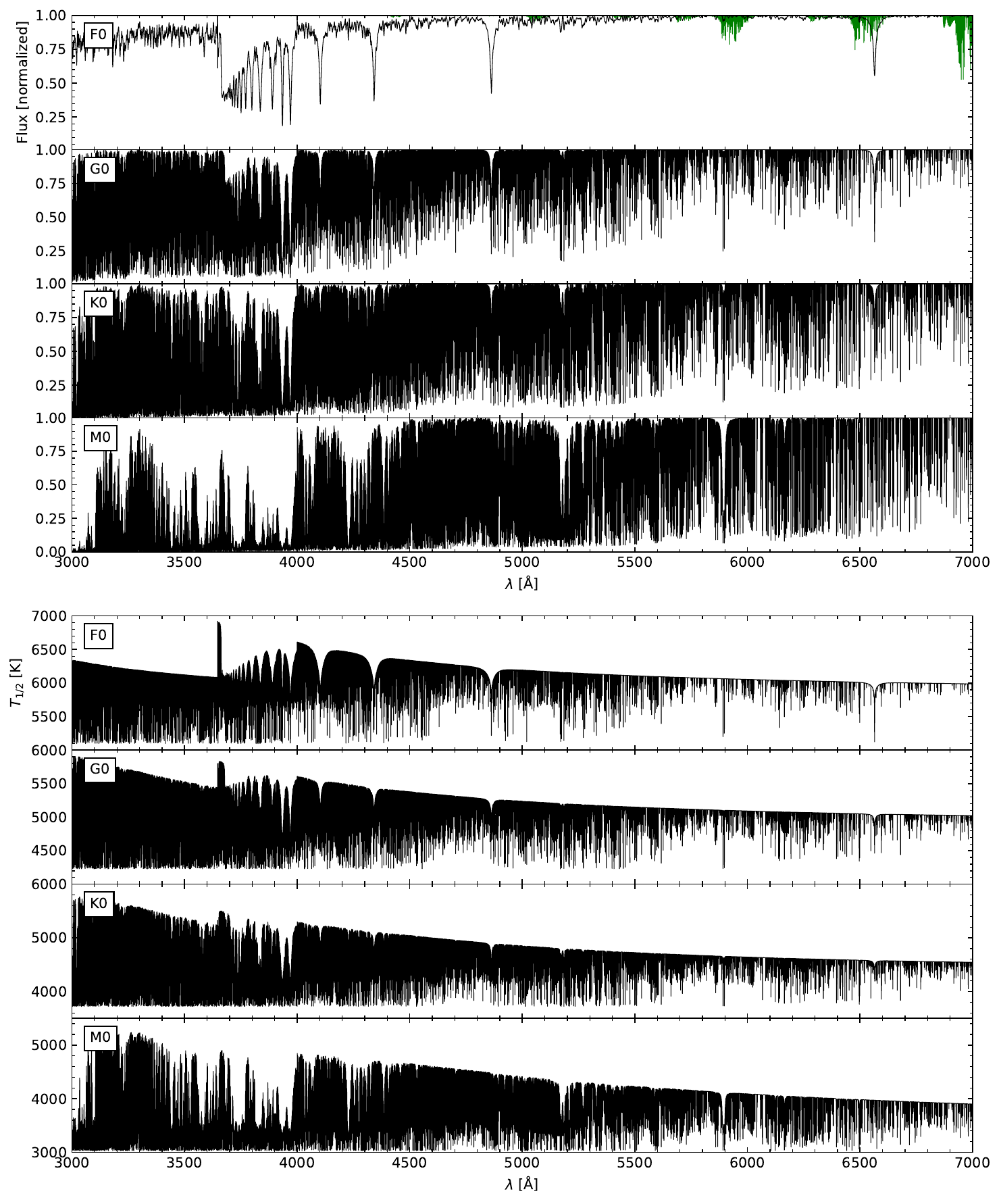}
    \caption{Sample of synthetic spectra. \textit{Upper panels}: Normalized flux spectra from the pre-computed grid between the wavelengths $\SI{3000}{}$--$\SI{7000}{\angstrom}$. Spectral types F0, G0, K0, and M0 are shown from top to bottom. In the first panel, the telluric spectrum is shown in green. \textit{Lower panels}: Same as the upper panels, but with the average formation temperature, $T_{1/2}$, instead of flux.}
    \label{Fig:A01}
\end{figure*}

\begin{figure*}[ht!]
    \includegraphics[width=\textwidth]{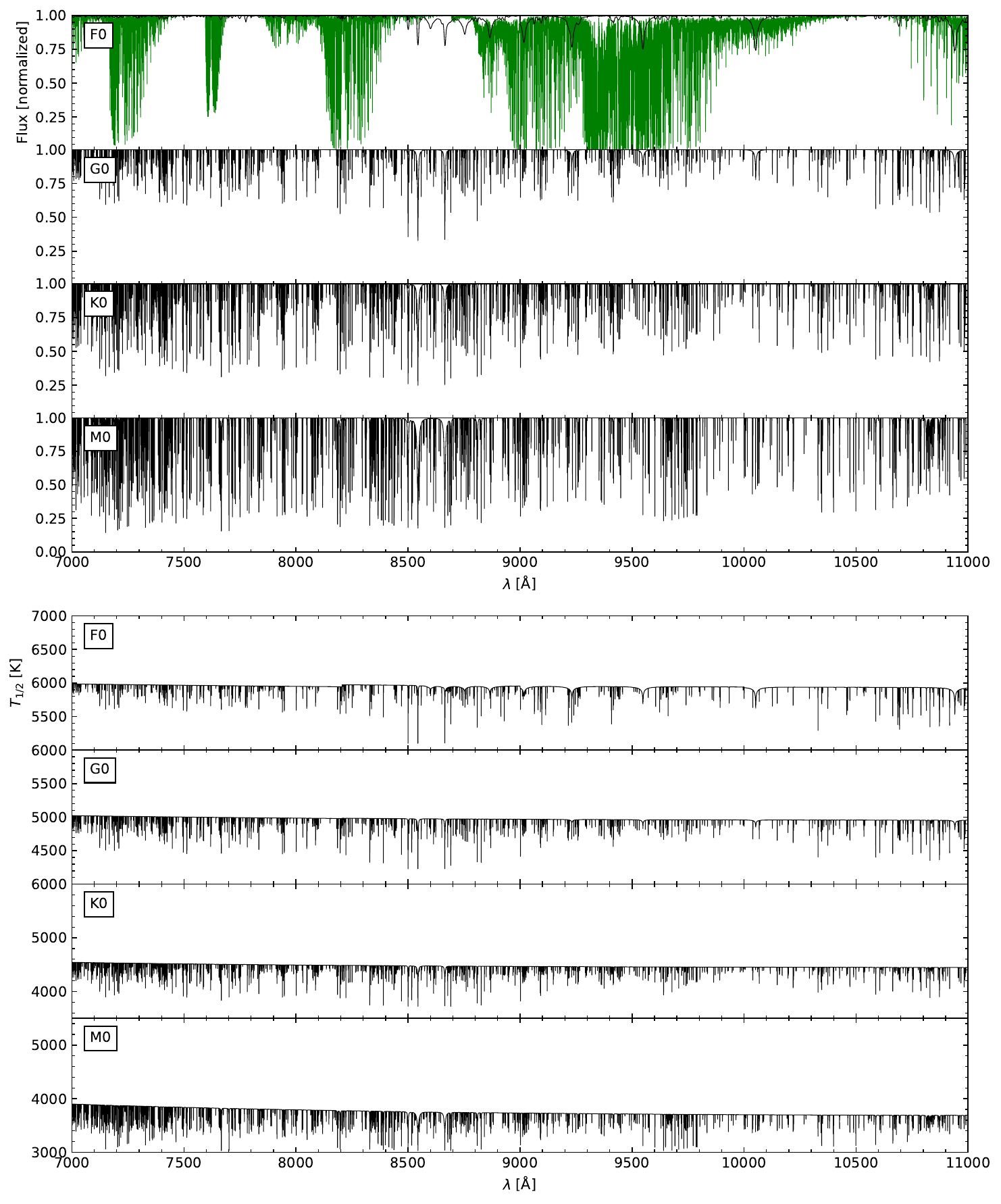}
    \caption{Same as Fig.~\ref{Fig:A01}, but for the wavelength range $\SI{7000}{}$--$\SI{11000}{\angstrom}$.}
    \label{Fig:A02}
\end{figure*}

\begin{figure*}[ht!]
    \includegraphics[width=\textwidth]{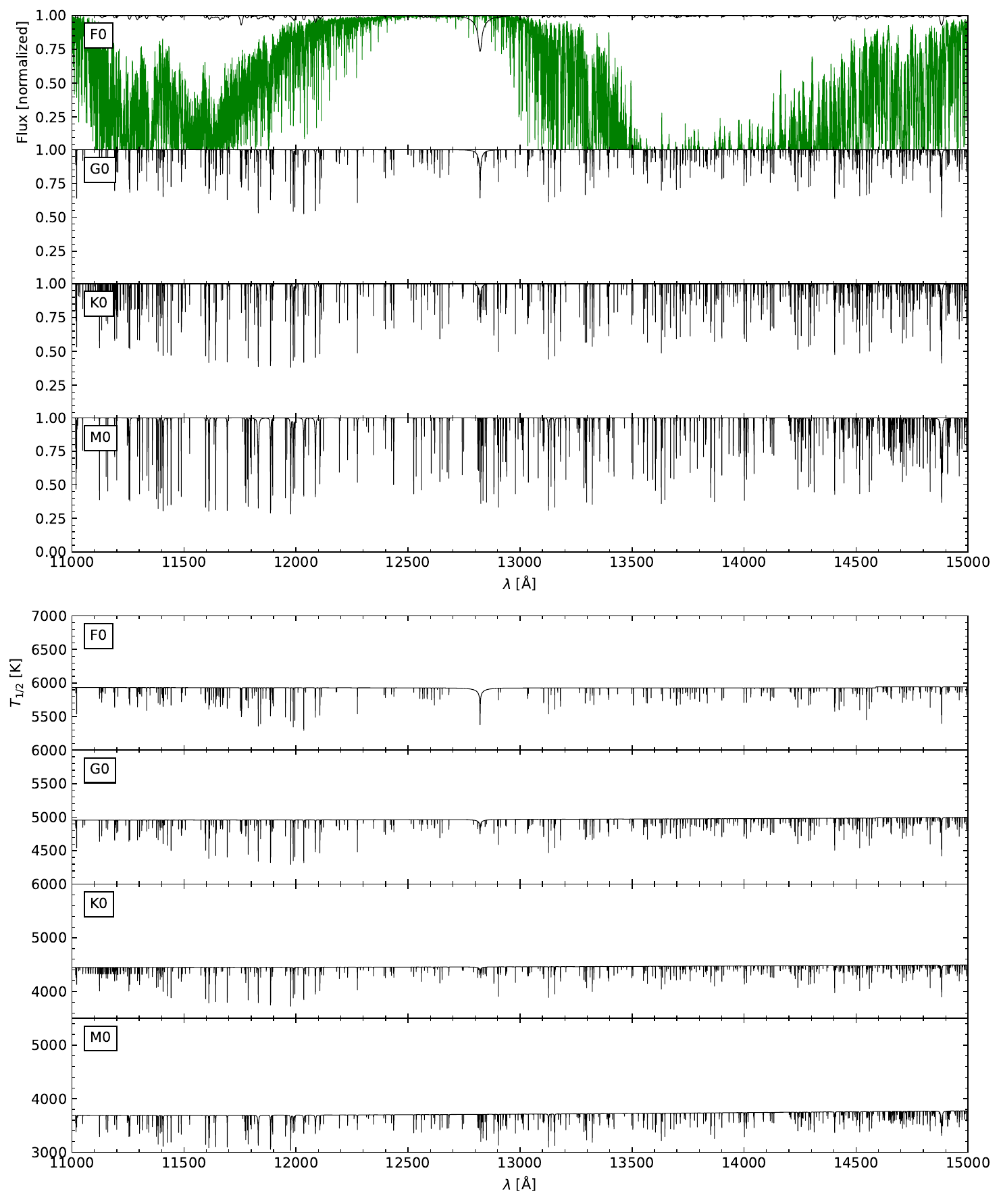}
    \caption{Same as Fig.~\ref{Fig:A01}, but for the wavelength range $\SI{11000}{}$--$\SI{15000}{\angstrom}$.}
    \label{Fig:A03}
\end{figure*}

\begin{figure*}[ht!]
    \includegraphics[width=\textwidth]{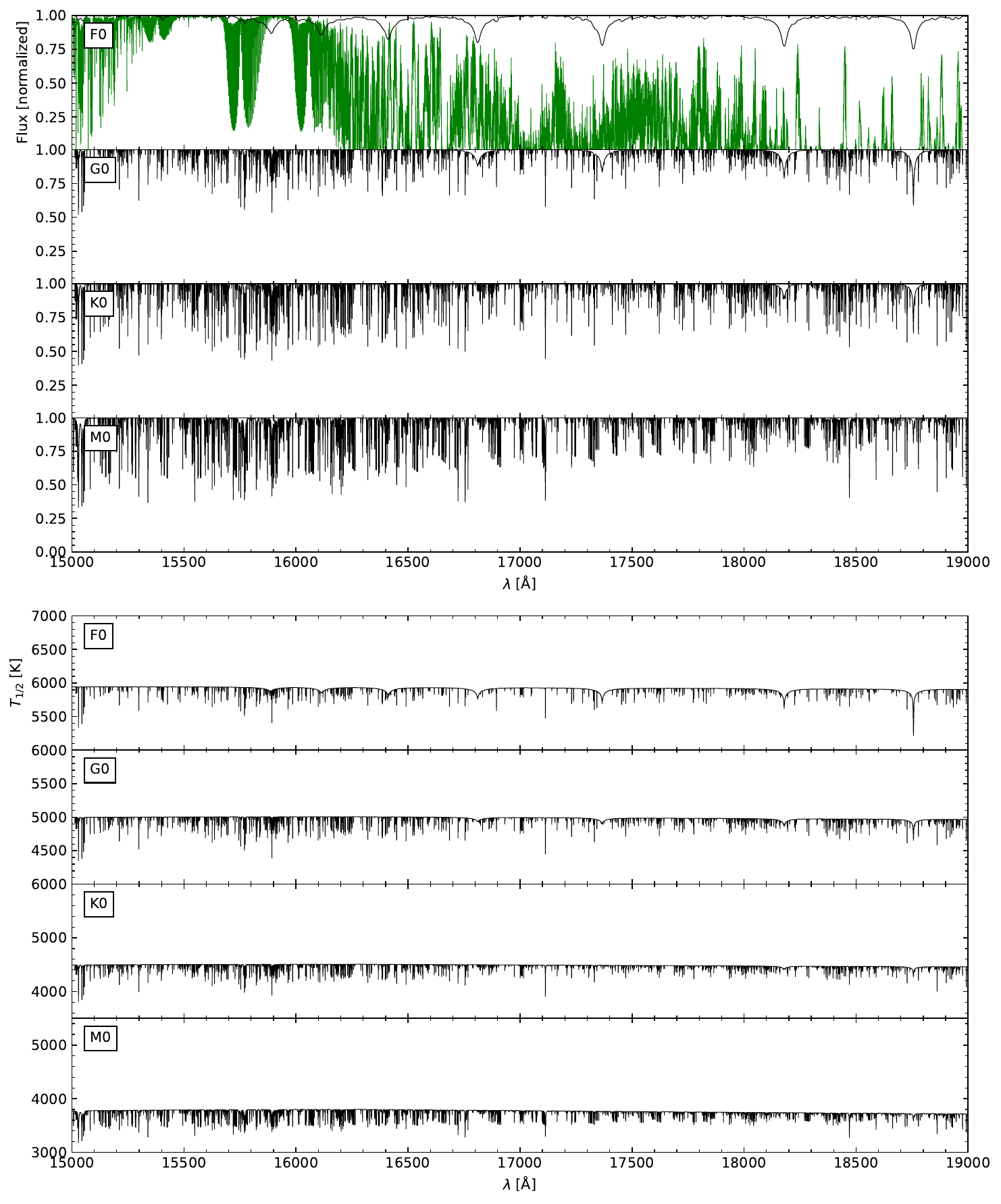}
    \caption{Same as Fig.~\ref{Fig:A01}, but for the wavelength range $\SI{15000}{}$--$\SI{19000}{\angstrom}$.}
    \label{Fig:A04}
\end{figure*}

\begin{figure*}[ht!]
    \includegraphics[width=\textwidth]{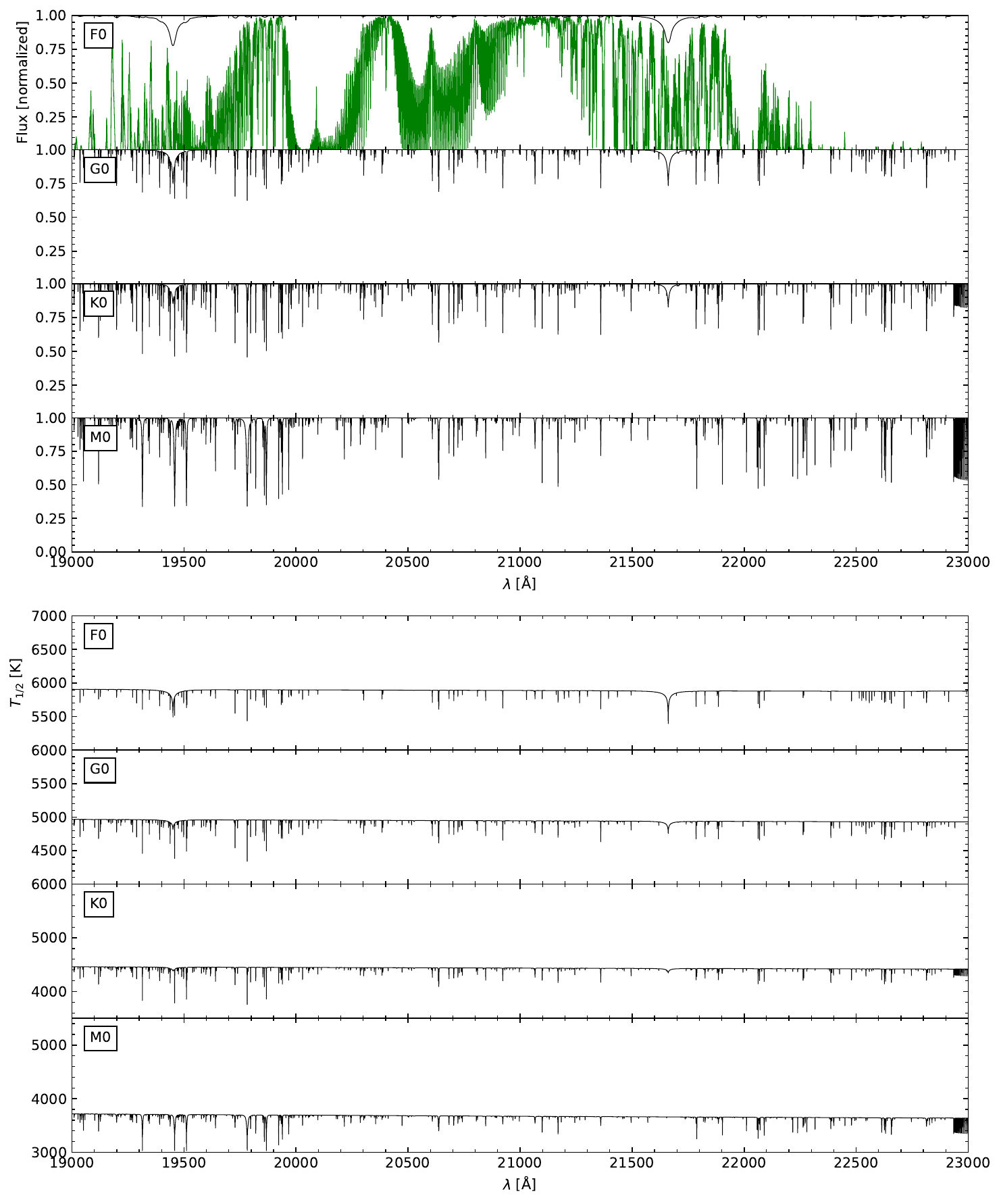}
    \caption{Same as Fig.~\ref{Fig:A01}, but for the wavelength range $\SI{19000}{}$--$\SI{23000}{\angstrom}$.}
    \label{Fig:A05}
\end{figure*}

\end{appendix}

\end{document}